\documentclass[12pt,draftclsnofoot,onecolumn]{IEEEtran}
\usepackage{graphicx}
\usepackage{epstopdf}
\usepackage{cite}
\usepackage{amsfonts}
\usepackage{amssymb}
\usepackage{amsmath}
\usepackage{mathrsfs}
\usepackage{bm}
\usepackage{amsmath}
\usepackage{mdwmath}
\usepackage{mdwtab}
\usepackage[ruled,vlined,linesnumbered]{algorithm2e} 
\usepackage{flushend}

\DeclareMathOperator*{\argmax}{arg\,max}
\newtheorem{remark}{Remark}
\newtheorem{theorem}{Theorem}
\newtheorem{lemma}{Lemma}
\newtheorem{prop}{Proposition}

\begin{document}
	\title{Joint User Scheduling and Beam Selection Optimization for Beam-Based Massive MIMO Downlinks}
	\author{Zhiyuan Jiang,~\IEEEmembership{Member,~IEEE}, Sheng Chen, Sheng Zhou,~\IEEEmembership{Member,~IEEE}, Zhisheng Niu,~\IEEEmembership{Fellow,~IEEE}
		\thanks{Z. Jiang, S. Chen, S. Zhou and Z. Niu are with Tsinghua National Laboratory for Information Science and Technology, Tsinghua University, Beijing 100084, China. Emails: \{zhiyuan@, chen-s16@mails., sheng.zhou@, niuzhs@\}tsinghua.edu.cn. 
			
			This work is sponsored in part by the Nature Science Foundation of China (No. 61701275, No. 91638204, No. 61571265, No. 61621091), the China Postdoctoral Science Foundation, and Intel Collaborative Research Institute for Mobile Networking and Computing. The corresponding author is Dr. Sheng Zhou.}}
	\maketitle
	\vspace{-10mm}
	\begin{abstract}
		In beam-based massive multiple-input multiple-output systems, signals are processed spatially in the radio-frequency (RF) front-end and thereby the number of RF chains can be reduced to save hardware cost, power consumptions and pilot overhead. Most existing work focuses on how to select, or design analog beams to achieve performance close to full digital systems. However, since beams are strongly correlated (directed) to certain users, the selection of beams and scheduling of users should be jointly considered. In this paper, we formulate the joint user scheduling and beam selection problem based on the Lyapunov-drift optimization framework and obtain the optimal scheduling policy in a closed-form. For reduced overhead and computational cost, the proposed scheduling schemes are based only upon statistical channel state information. Towards this end asymptotic expressions of the downlink broadcast channel capacity are derived. To address the weighted sum rate maximization problem in the Lyapunov optimization, an algorithm based on block coordinated update is proposed and proved to converge to the optimum of the relaxed problem. To further reduce the complexity, an incremental greedy scheduling algorithm is also proposed, whose performance is proved to be bounded within a constant multiplicative factor. Simulation results based on widely-used spatial channel models are given. It is shown that the proposed schemes are close to optimal, and outperform several state-of-the-art schemes.
	\end{abstract}
	\section{Introduction}
	\label{sec_intro}
	In massive multiple-input multiple-output (MIMO) based wireless communication systems, the spectral and radiated energy efficiency can be both boosted by the deployment of massive number of antennas \cite{Marzetta10}. Moreover, the high beamforming gain of a massive antenna array is the main enabler for millimeter-wave systems against high pathloss. Therefore, it is extremely important to design high-performance, efficient and practical transmission strategy in massive MIMO systems for the emerging $5$G cellular system. 
	
	Under the assumption that full digital signal processing is performed at the base station (BS) side with massive antenna arrays, the system performance has been widely investigated, e.g., in \cite{Marzetta10,Rusek12,larsson14}. However, it is widely accepted that full digital signal processing implementation encounters very severe challenges in practice, on account of the following impediments.
	
	\textbf{Radio-frequency (RF) chain hardware cost and power consumptions}. Full digital signal processing requires that all antennas can be digitally controlled from baseband. Hence, one dedicated RF chain, including e.g., low-noise amplifier, analog-digital-converter (ADC), power amplifier and etc., is needed for each antenna. In massive MIMO systems, not only is this requirement entails a dramatic increase in the deployment cost of the system, but also that the power consumption would be driven up to a prohibitive level. As indicated in the previous work \cite{heath16} \cite{han15}, concretely, a BS with $256$ RF chains consumes about $10$ times the power (only the RF chains) as compared with an entire current long-term-evolution (LTE) BS. 
	
	\textbf{Baseband signal processing complexity.} The spatial baseband processing includes multiple kinds of matrix operations, such as inversions and singular-value-decompositions (SVDs) whose complexity scales with $M^3$ where $M$ is the number of antenna elements for full digital processing. Moreover, these extremely demanding matrix operations are required to be executed very frequently (once every $1$~{ms} for spatial precoding in LTE systems). This is very challenging to the design of baseband processing units, both in terms of chip costs and power consumption. 
	
	\textbf{System specific limitations.} Aside from the first two challenges, there are some other practical considerations which are system-specific. For example, the fronthaul interface in cloud radio access networks (C-RAN) poses a serious limitation in the number of data streams that can be transmitted between the remote-radio-units (RRUs) and the baseband units (BBUs). Considerable amounts of work has been dedicated to the signal spatial compression in C-RAN \cite{jiang_icc17}. Moreover, the channel state information (CSI) acquisition overhead in frequency-division-duplexing (FDD) system scales with the number of digitally controllable antennas. It constitutes a major bottleneck in realizing the massive MIMO gain in FDD systems.
	
	In view of these challenges, architectures with low RF- and processing-complexity have been proposed extensively, e.g., in \cite{alkh14,molisch16,gao16,brady13,Jiang14,7,jiang17}. The existing literature can be divided into three categories. The first is \emph{hybrid beamforming}, which adopts an RF front end with an analog beamforming module such that the number of RF chains is significantly reduced \cite{molisch16}. Although the analog beamforming module is usually composed of phase shifters with constant-amplitude beamforming weights to save hardware cost, the high-speed phase shifters, whose quantity is the same with the number of antenna elements, pose a drastic challenge to the cost of RF front ends. In this regard, the recently proposed \emph{beamspace MIMO} architecture \cite{brady13} adopts a lens antenna array which acts analogously like a lens focusing on light beams from different directions. It transforms the signal into the angular domain and thus reduces the number of RF chains due to signal angular sparsity. Since it does not require any phase shifters, the total cost is reduced, and therefore it is considered to be one of the candidate solutions to the 5G millimeter-wave massive MIMO systems. The other approach is based on digital beamforming which involves \emph{multi-layer signal processing} \cite{Adhikary13,heath16,Jiang14}. Although the number of RF chains is not reduced, the processing complexity and pilot overhead problems are partly solved. 
	\begin{figure}[!t]
		\centering
		\includegraphics[width=0.9\textwidth]{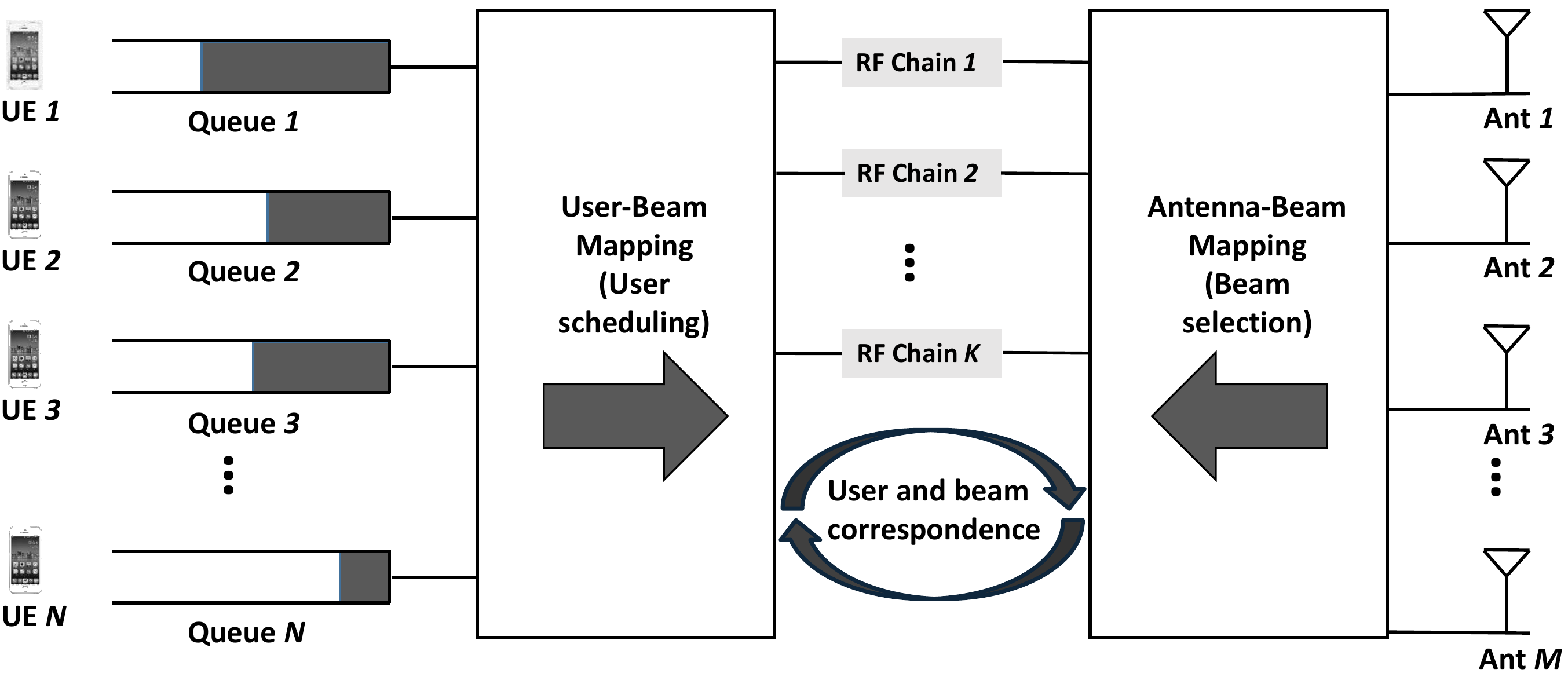}
		\caption{An illustration of the beam-based massive MIMO systems where user scheduling and beam selection are correlated.}
		\label{Fig_arch}
	\end{figure}
	
	In essence, all the aforementioned solutions aim at providing comparable performance as full digital processing systems with limited number of RF chains and reduced complexity in massive MIMO systems. Based on Fig. \ref{Fig_arch}, the current literature mainly focuses on the right side of the figure, i.e., the antenna to beam mapping and beam selection schemes, which leverages the angular domain power sparsity of the channel to transform the signals from the antenna domain to the beam domain. Towards this end, beam sweeping and steering methods in the hybrid beamforming architecture can be used to capture the signal direction \cite{molisch04}. Other methods adopt joint analog and digital precoding design \cite{gao16} \cite{Ayach14}. In beamspace MIMO systems, the lens antenna array can be regarded as a directional beamforming module with low cost. On the other hand, the left side of the Fig. \ref{Fig_arch} which represents user-beam mapping, is scantly treated. The user-beam mapping essentially deals with \textbf{user scheduling in the beam domain}. Unlike the previous user scheduling related work, e.g., in \cite{Yoo2006, shirani10, kim04}, the user scheduling problem in the beam domain is tangled with the beam selection. In reality, due to the angular sparsity of the massive MIMO channel \cite{bas17}, the beams, which represent the signal directions, are strongly related to the users, in the sense that each beam usually contains signals of very few (possibly one) users. Therefore, the user scheduling and beam selection have to be jointly considered to avoid possible performance degradation due to user-beam mismatch. 
	
	The channel state information (CSI) is of vital importance to the system. The CSI can be categorized as instantaneous CSI and statistical CSI. It is worthwhile to emphasize the specific CSI usage at each stage (time scale) of the beam-based massive MIMO transmissions since most existing work ignores this and assumes instantaneous CSI is always available \cite{gao16_bs} \cite{ama15} . We propose that beam-based downlink scheduling should be performed \emph{only} based on the \textbf{statistical CSI}. The reason is two-fold. From an implementation perspective, the statistical CSI is much easier to obtain than instantaneous CSI, attributing to the fact that statistical CSI can be obtained with much lower cost because a) it can be estimated without dedicated pilots \cite{liang01}; b) it varies at a lower speed (in the order of $1$~second to $10$~seconds) compared with instantaneous CSI (in the order of $1$~{ms} to $100$~{ms}) \cite{Adhikary14}. Moreover, in C-RAN systems, the beamforming module is integrated with the remote radio heads (RRHs) and hence limited computation capability is expected \cite{che15} which prevents us from using complicated channel estimation schemes. On the other hand, it is also theoretically possible to only rely on statistical CSI in the user scheduling and beam selection phase, since beams are essentially long-term statistics. Furthermore, the statistical CSI can be obtained efficiently with a limited number of RF chains based on, e.g., compressive sensing based channel estimation schemes \cite{alkh14} \cite{gao17_sd}.
	
	In this paper, we aim to address the user and beam joint scheduling problem in beam-based massive MIMO downlinks. The contributions include:
	\begin{itemize}
		\item 
		We formulate the problem based on the Lyapunov-drift optimization framework. An optimal scheduling policy is proposed thereby to achieve optimum utilities. The optimality proof is given which shows the achieved utility is arbitrarily close to the optimum.
		\item
		To address the queue weighted sum rate maximization problem arisen in optimizing the Lyapunov-drift, which is a mixed integer programming (MIP) problem, the block-coordinated-update-based (BCU-based) algorithm which deals with the continuous convex relaxation of the MIP problem is proposed. In order to implement the algorithm based on statistical CSI, a deterministic equivalence of the downlink broadcast channel capacity in the large antenna array regime is derived, depending only on statistical CSI. The BCU-based algorithm is proved to converge to the global optimum of the relaxed problem. An iterative water-filling based approach is also proposed to reduce the number of iterations.
		\item
		Furthermore, a low-complexity incremental greedy algorithm is proposed. We prove that the greedy algorithm can achieve near-optimal performance, within a multiplicative factor due to the submodular property of the problem.
		\item
		By simulations, it is shown that the proposed algorithms outperforms several state-of-the-art beam selection schemes. Moreover, since it is based on statistical CSI, the frequency of executing the algorithm is significantly reduced, making it more preferable in practice.
	\end{itemize} 
	
	\subsection{Related Work}
	The proposed joint user and scheduling schemes are related to the beam selection problem in beamspace MIMO systems \cite{gao16_bs,ama15,jiang17_apcc,wang16,sun15,you17}, or more generally antenna selection problem in MIMO systems \cite{molisch04mag}. However, the considered joint scheduling problem in beam domain is unique, in the sense that the beam magnitudes are strongly correlated with users. The beam-user scheduling problem is also considered in a switched-beam based massive MIMO system in \cite{wang16}, where one pre-defined (\emph{fixed}) beam is associated with one user. In \cite{sun15}, a greedy joint scheduling of beams of users is proposed and we will compare our results with it in the simulations. Concerning the literature related to the mathematical treatment adopted in the paper, the Lyapunov-drift optimization framework is attributed to the pioneer work by Neely \cite{neely10}. The large system deterministic equivalence to derive the downlink channel capacity is related to the celebrated random matrix theory \cite{edelman05}. Furthermore, in the algorithm design, the BCU technique dates back to multi-convex optimization, e.g., in \cite{warga63}, and the approximation factor of the greedy algorithm is related to the submodular set function optimization problem as in \cite{nem78}.
	\subsection{Paper Organizations and Notations}
	The remainder of the paper is organized as follow. In Section \ref{sec_sm}, the system model and channel model are presented, and the problem is formulated. In Section \ref{sec_lya}, the Lyapunov-drift approach is used to design an optimal scheduling policy. In Section \ref{sec_bcu}, a BCU-based scheduling algorithm is described to address the queue weighted sum rate maximization problem. In Section \ref{sec_igs}, a low-complexity greedy algorithm is presented. The simulation results are conveyed in Section \ref{sec_sr}. Finally, we conclude our work in Section \ref{sec_cl}.
	
	Throughout the paper, we use boldface uppercase letters, boldface lowercase letters and lowercase letters to designate matrices, column vectors and scalars, respectively. $\bm{X}^T$ and $\bm{X}^\dag$ denotes the transpose and complex conjugate transpose of matrix $\bm{X}$, respectively. $X_{i,j}$ and $x_i$ denotes the $(i,j)$-th entry and $i$-th element of matrix $\bm{X}$ and vector $\bm{x}$, respectively. $\textnormal{tr}(\bm{X})$ denotes the trace of matrix $\bm{X}$. Denote by $\mathbb{E}(\cdot)$ as the expectation operation. Denote by $\bm{I}_N$ as the $N$ dimensional identity matrix. The logarithm $\log(x)$ denotes the binary logarithm. 
	
	\section{System Model and Problem Formulation}
	\label{sec_sm}
	\subsection{Signal Model}
	The single-cell system downlink is considered in this paper, where a BS with $M$ co-located antennas transmits to $N_\textrm{t}$ single-antenna users.\footnote{The proposed schemes can be straightforwardly extended to multiple-antenna-user case by treating multiple antennas of a user as multiple users with an identical channel correlation matrix \cite{liang01} and setting the backlog pressure $Q_n(t)$ in \textbf{P3} to be the same for these antennas.} The BS has $K$ RF chains ($K \le M$). Assuming narrow-band and time-invariant channels\footnote{Wideband channels can be decomposed to a set of parallel narrow-band channels by, e.g., orthogonal-frequency-division-multiplexing (OFDM) modulations. The time-invariant channel assumption essentially deals with data transmission inside the channel coherence time (or block length in the block fading channel model).}, the receive signal of user-$n$ is, , 
	\begin{equation}
	\label{signal_model}
	y_n =  \bm{h}_n^\dag \bm{x}  + n_n,
	\end{equation}
	where $\bm{h}_n$ is an $M$-dimensional channel vector, $\bm{x}$ is the downlink transmit signals, and $n_n$ denotes the i.i.d. Gaussian additive noise with unit variances. The downlink channel matrix is denoted by $\bm{H} = \left[\bm{h}_1,\bm{h}_2,...,\bm{h}_{N_\textrm{t}}\right]^\dag$. The transmit signal after beamforming can be written as
	\begin{equation}
	\label{precoding}
	\bm{x} =  \bm{B}_\textrm{a} \bm{s},
	\end{equation}
	where $\bm{s}$ denotes the $K$-dimensional digitally precoded data symbols for the scheduled, i.e., spatial-multiplexed $N_\textrm{s}$ users ($N_\textrm{s} \le K$ such that a linear digital precoder such as zero-forcing precoder can eliminate the inter-user-interference,\footnote{In the simulations, we assume $K=N_\textrm{s}$ and hence $K \le N_\textrm{t}$.} and obviously $N_\textrm{s} \le N_\textrm{t}$). The RF (analog) beamforming, which can be realized by the lens antenna array and beam selection in beamspace MIMO or general analog beamforming in hybrid beamforming architectures, is denoted by $\bm{B}_\textrm{a}$ with dimension $M \times K$. On account of the analog beamforming, the effective channel observed from baseband is
	\begin{equation}
	\label{He}
	\bm{\bar H} = \bm{H} \bm{B}_\textrm{a},
	\end{equation}
	where the effective channel vector corresponding to user-$n$ is denoted by $ \bm{\bar h}_n$ and $\bm{\bar H} = \left[\bm{\bar h}_1,...,\bm{\bar h}_{N_\textrm{t}}\right]^\dag$. 
	The RF beamforming considered in the paper is a widely adopted directional beamforming scheme\footnote{Note that the RF beamforming in this paper can be readily generalized to arbitrary beam pattern, e.g., the eigenvector-based beam pattern in \cite{Adhikary13}, by replacing the DFT-based beamforming matrix $B_\textrm{DFT}$ with the desired beamforming matrix. Also note that although directional beamforming is adopted, the proposed schemes can adapt to the channel variation more flexibly compared with traditional directional antenna based systems, by updating the channel statistics estimations and adjusting beam patterns.}, and hence
	\begin{equation}
	\label{ba}
	\bm{B}_{\textrm{a}} = \bm{B}_\textrm{DFT}\bm{\Sigma}_\textrm{b},
	\end{equation}
	where $\bm{B}_\textrm{DFT}$ is the equivalent discrete-Fourier-transform (DFT) matrix (or Kronecker product of DFT matrices for uniform planar antenna arrays (UPAs)). The beam selection decision is denoted by the diagonal matrix $\bm{\Sigma}_\textrm{b}$ whose entries are binary, i.e., $(\bm{\Sigma}_\textrm{b})_{i,i} \in \{0,1\},\,\forall i$. 
	
	\subsection{Channel Model}
	\label{sec.cm}
	Based on a geometry-based channel model \cite{molisch04_gscm}, the channel vector of the $n$-th user can be written as
	\begin{equation}
	\label{ray_rep}
	\bm{h}_{n} = \sqrt{\frac{M}{L_{n}}}\sum\limits_{l=1}^{L_{n}}\beta_l^{(n)} \bm{\alpha}\left(\theta_l^{(n)},\psi_l^{(n)}\right),
	\end{equation}
	where $L_{n}$ denotes the total number of multi-path components (MPCs) in the propagation medium including line-of-sight (LoS) and none-line-of-sight (NLoS) MPCs. The amplitude of each MPC is denoted by $\beta_l^{(n)}$, and $\theta_l^{(n)}$ and $\psi_l^{(n)}$ denote the azimuth and elevation angles of the $l$-th arriving MPC, respectively. Thereby, the steering vector for one MPC is given by (assuming UPA whereas one-dimensional uniform-linear-array (ULA) can be regarded as a special case)
	\begin{IEEEeqnarray}{rCl}
		\label{array_res}
		\bm{\alpha}^{\textrm{UPA}}\left(\theta_l^{(n)},\psi_l^{(n)}\right) = \frac{1}{\sqrt{M}}\left[1,...,e^{-j2\pi\left(m\lambda_h\sin{\theta_l^{(n)}}\cos{\psi_l^{(n)}}+n\lambda_v\cos{\theta_l^{(n)}}\sin{\psi_l^{(n)}}\right)}\right.,..., \nonumber \\
		\left.e^{-j2\pi\left((H-1)\lambda_h\sin{\theta_l^{(n)}}\cos{\psi_l^{(n)}}+(V-1)\lambda_v\cos{\theta_l^{(n)}}\sin{\psi_l^{(n)}}\right)}\right]^T,
	\end{IEEEeqnarray}
	where $H$ and $V$ denote the number of columns and rows in the UPA, respectively, and $\lambda_h$ and $\lambda_v$ are the antenna spacing in the horizontal and vertical domains, respectively. The order of elements in the steering vector is mapped to the indexing order of antennas in the UPA. The physical meaning of the steering vector and channel representation in \eqref{ray_rep} is that for an MPC with direction-of-arrival (DoA) $\left(\theta_l^{(n)},\psi_l^{(n)}\right)$, the array response is given by \eqref{array_res}. Summing up all the contributing MPCs, we obtain the compound channel representation in \eqref{ray_rep}. Based on the channel model, the RF beamforming in \eqref{ba} can take advantage of the limited number of MPCs as compared with the number of antennas, and only selects a subset of the beams to attain equivalent performance (signal power) with a smaller number of RF chains. 
	
	However, the beam selection cannot be isolated from the user scheduling problem. Apart from the reasons given in Section \ref{sec_intro}, from a throughput perspective, different users have different transmission needs resulting from traffic demand or fairness considerations. Therefore, the user scheduling and beam-domain CSI should also be jointly considered. The joint beam-domain massive MIMO downlink scheduling problem is formulated as follows.
	
	\subsection{Problem Formulations}
	The long-time average rate of user $n$ is denoted by $\bar R_n$, and the instantaneous rate of user $n$ at time $t$ is denoted by $R_n(\bm{H}(t),\pi(t))$, given the channel coefficients $\bm{H}(t)$ and control policy (user scheduling and beam selection as far as the paper is concerned) $\pi(t)$. Note that this does not mean the scheduling decision relies on the availability of instantaneous CSI, as in Proposition \ref{prop1} an deterministic equivalence of the rate expression will be derived which is based solely upon statistical CSI. Based on ergodicity, ${{\bar R}_n} = \mathbb{E}\{ {R_n}(\bm{H},{\pi})\}$, $\forall n \in \{1,...,N_\textrm{t}\}$, where the expectation is taken over channel coefficients $\bm{H}(t)$ and possibly ${\pi}(t)$ when a stochastic control policy is considered. The achievable ergodic rate region can be characterized as
	\begin{equation}
	\label{EgrateRegion}
	\mathcal{R }= \textrm{coh}\bigcup\limits_{{\pi } \in \mathcal{X}} {\left\{ \bm{\bar R}:0 \le {{\bar R}_n} \le \mathbb{E}\left[{R_n}(\bm{H},{\pi })\right]\right\} },
	\end{equation}
	where $\mathcal{R }$ is a $N_\textrm{t}$-dimensional region, ${{\bar R}_n} $ is its $n$-th component, and ``coh'' denotes the closure of a convex hull. The set of all feasible scheduling policies is denoted by $\mathcal{X}$. The utility maximization problem is formulated as
	\begin{flalign}
		\label{p1}
		\textbf{P1:}&&\mathop{\textrm{maximize}}\limits_{\bm{\Sigma}_u,\bm{\Sigma}_b}  U\left({{\bar{ \bm{R}}}}\right),\,\textrm{subject to } \, {\bar{ \bm{R}}} \in \mathcal{R},&&
	\end{flalign}
	where $\bm{\Sigma}_\textrm{u}$ is a diagonal matrix denoting user scheduling decision at time $t$, i.e., $s_i  \overset{\Delta}{=} \left( {\bm{\Sigma}_\textrm{u}}\right)_{i,i} \in \{0,1\}$, and 
	$b_i  \overset{\Delta}{=} \left({\bm{\Sigma}_\textrm{b}}\right)_{i,i}$ denotes the beam selection decision. The network utility function $U\left({\bar{ \bm{R}}}\right)$ is defined as a function of the long-time average rate for each user, e.g., 
	\begin{equation}
	\label{sr_utility}
	U_\textrm{sum}\left({\bar{ \bm{R}}}\right) = \sum_{n=1}^{N_\textrm{t}} {\bar R}_n
	\end{equation}
	for sum rate maximization, 
	\begin{equation}
	\label{pfs_utility}
	U_\textrm{pfs}\left(\bar{ \bm{R}}\right) = \sum_{n=1}^{N_\textrm{t}} \log ({\bar R}_n + c_n)
	\end{equation}
	for proportional-fairness scheduling (PFS) \cite{kush04}, where $c_n$'s are non-negative constants to regularize the logarithm expressions, and typical value is $c_n=0$, $\forall n \in \{1,...,N_\textrm{t}\}$ for exact PFS or $c_n=1$, $\forall n \in \{1,...,N_\textrm{t}\}$ to ensure positive objective function value which is a mathematical convenience. Basic properties of the utility function $U\left(\bar{ \bm{R}}\right)$ are required, e.g., concavity and monotonicity \cite{neely08}, over the rate vector $(R_1,...,R_{N_\textrm{t}})$.
	
	\section{Optimal Beam-Based Joint Scheduling Policy}
	\label{sec_lya}
	To solve \textbf{P1}, it is found that two severe challenges exist. First, the ergodic capacity region $\mathcal{R}$ does not yield a closed-form expression. The work in \cite{Weingarten06} \cite{Wei06} characterizes the broadcast channel (BC) capacity region and a duality between BC and multiple-access-channel (MAC) in the sense of both capacity region and outage probability is found. Moreover, an iterative water-filling algorithm is proposed to calculate it, given the \emph{instantaneous} channel coefficients. Nevertheless, no closed-form expressions are available except for capacity bounds \cite{Jiang12}.\footnote{The broadcast channel capacity formula is adopted as the optimization objective in this paper due to its better generality compared with, e.g., achievable rates based on linear precoding schemes. It is also because that non-linear downlink transmission schemes, e.g., non-orthogonal multiple access schemes, are attracting more and more attention recently.} Secondly, the scheduling and beam selection decisions should be made \emph{dynamically} to match the channel variations and user traffic in time. To address these issues, we seek to leverage a powerful tool of Lyapunov-drift optimization which is shown to have superior performance compared to static solutions \cite{Jang02} with simple decision structures (max-weight structure \cite{geo06}); thereby, \textbf{P1} is decomposed into \textbf{P2} and \textbf{P3} described in the following. 
	
	Essentially, \textbf{P1} is a time-average network utility maximization problem which is hard to solve directly. The Lyapunov-drift approach decomposes the time-average optimization into optimization in each scheduling step; and the resultant sub-problems are formulated as \textbf{P2} and \textbf{P3}. By applying the solutions to \textbf{P2} and \textbf{P3} at each scheduling step, the time-average network utility can be optimized.  
	
	\subsection{Lyapunov-Drift Based Network Utility Maximization}
	To maximize the network utility function in \eqref{p1}, the transmission need of each user, which is determined by the transmission history and utility function, is represented by a set of virtual queues. The arrival process is designed to reflect the transmission urgency of each user and a max-weight algorithm is applied to stabilize the queues whenever possible. Specifically, let $Q_n(t)$ denote the virtual queue length in bits of user $n$ at the beginning of $t$-th scheduling step, let $a_n(t)$ denote the number of (virtual) arrival bits which are optimization variables for utility maximization, and let $\mu_n(t)$ denote the allocated number of service bits to queue-$n$, which equals the allocated number of service bits between scheduling steps. The queuing dynamics are written as
	\begin{equation}
	\label{QueueD}
	{Q_n}(t+1) = {Q_n}(t) - {\tilde \mu _n}(t) + {a_n}(t),
	\end{equation}
	where ${\mu _n}(t) = \sum_{\tau=1}^TR_n(\bm{H}(\tau),\pi(\tau))$, the number of channel uses between scheduling steps is denoted by $T$, and ${{\tilde \mu }_n}(t) = \min \{ {Q_n}(t),{\mu _n}(t)\}$ denotes the number of actual service bits, considering the circumstances that sometimes the queue is emptied given the amount of allocated service bits. Notice that the queues here are created virtually to facilitate the utility maximization and thus they are not real traffic patterns. In Section \ref{sec_sr} (Fig. \ref{Fig_traffic}), we extend to stochastic real traffic scenarios in simulations. The optimal beam-based downlink scheduling policy at a given scheduling time $t$, i.e., a dynamic policy which achieves the solution to \eqref{p1}, can be described as below.
	
	\textbf{Admission control}: For virtual queue $\bm{Q}(t) = [Q_1(t),...,Q_{N_\textrm{t}}(t)]$, let the number of arrival bits, i.e., $\bm{a}(t)$, be the solution of 
	\begin{flalign}
		\label{p2}
		\textbf{P2:}&&\mathop{\textrm{maximize}}\limits_{\bm{a}(t)} \, V U\left({\bm{a}(t)}\right) - \bm{a}(t)^T\bm{Q}(t), \,\,\,\,\,\, \textrm{subject to } 0 \le a_n(t) \le A_\textrm{max},\,\forall n \in \{1,...,N_\textrm{t}\},&&
	\end{flalign}
	where $V$ and $A_\textrm{max}$ are pre-defined constants\footnote{For Typical values, $V$ and $A_{\textrm{max}}$ can be approximately $100$-fold of the service rate.}.
	
	\textbf{Scheduling}: Given the arrival process determined above, the service, i.e., the joint scheduling decisions, is based on the solution of the following problem:
	\begin{flalign}
		\label{P2}
		\textbf{P3:}&&\mathop{\textrm{maximize}}\limits_{\bm{\Sigma}_\textrm{u},\bm{\Sigma}_\textrm{b},\bm{p}}  \,\,& \sum_{n=1}^{N_\textrm{t}} \left[Q_n(t)  s_n\sum_{\tau=1}^T R_n\left(\bm{H}(\tau),\bm{\Sigma}_\textrm{u},\bm{\Sigma}_\textrm{b},\bm{p}\right)\right] &&\\
		\label{p_const}
		&&\textrm{s.t.,}\,\,  & \sum_{n=1}^{N_\textrm{t}} p_n \le P,&&\\
		&&& \sum_{i=1}^{N_\textrm{t}} s_{i} = N_\textrm{s},\,\sum_{i=1}^M b_i = K,\,s_i \in \{0,1\},\,b_i \in \{0,1\},&&
	\end{flalign}
	where $\bm{p} = [p_1,...,p_{N_\textrm{t}}]$ denotes the transmit power corresponding to $N_\textrm{t}$ user data streams and hence $P$ in \eqref{p_const} is the sum power constraint. The scheduling decisions are denoted by binary variables $s_i$ and $b_i$. $\bm{\Sigma}_\textrm{b}$ and $\bm{\Sigma}_\textrm{u}$ are diagonal matrices consisting of $s_i$ and $b_i$, respectively. The downlink instantaneous transmission 
	rate $R_n\left(\bm{H}(t),\bm{\Sigma}_\textrm{u},\bm{\Sigma}_\textrm{b},\bm{p}\right)$ is a function of the downlink transmit power allocation, channel coefficients, and scheduling decisions. The departure from the $n$-th virtual queue is ${\mu _n}(t) = s_n\sum_{\tau=1}^T R_n\left(\bm{H}(\tau),\bm{\Sigma}_\textrm{u},\bm{\Sigma}_\textrm{b},\bm{p}\right)$.
	
	It is observed that the admission control problem \textbf{P2} is a convex problem and hence is easy to solve. For instance with PFS utility, the optimal admission control is given by
	\begin{equation}
	a_n^*(t) = \min\left\{\frac{V}{Q_n(t)},A_\textrm{max}\right\},\, n \in \{1,...,N_\textrm{t}\}.
	\end{equation}
	However, the problem \textbf{P3} is an MIP problem, which is NP-complete \cite{karp72}. Before diving into details on solving \textbf{P3} in the following sections, we assume the optimal solutions to both problems are obtained for the moment, which is denoted by $\pi^*$. The optimality of the algorithm is established in the following theorem. 
	\begin{theorem}
		\label{thm1}
		Denote 
		\begin{equation}
		\bar{\bm{R}}^* = \mathop{\argmax}\limits_{ {\bar{\bm{R}}} \in \mathcal{R }} U\left({{\bar{ \bm{R}}}}\right).
		\end{equation}
		Suppose the transmission rate is bounded, i.e., $0 \le \bar{R}_n \le R_\textrm{max},\,\forall n \in \{1,...,N_\textrm{t}\}$, the utility function $U(\cdot)$ is concave and entry-wise non-decreasing, and bounded on $\left[0,R_\textrm{max}\right]$. The channel coefficients $\bm{H}(t)$ are i.i.d. over different scheduling periods, then based on the scheduling algorithm resulting from \textbf{P2} and \textbf{P3}, the following conditions are met.
		\begin{IEEEeqnarray}{rCl}
			\label{opt}
			\mathop{\liminf}\limits_{\tau \to \infty} U\left(\frac{1}{\tau} \sum_{t=0}^{\tau-1} \mathbb{E}[\bm{R}(t)]\right) &\ge& U\left(\bar{\bm{R}}^*\right) - C/V,\\
			\mathop{\lim}\limits_{\tau \to \infty} \frac{\mathbb{E}\left[Q_n(\tau)\right]}{\tau} &=& 0,\,\forall n.
		\end{IEEEeqnarray}
	\end{theorem}
	\begin{IEEEproof}
		See Appendix \ref{app_1}.
	\end{IEEEproof}
	\begin{remark}
		Theorem \ref{thm1} reveals that the utility function of the time-averaged transmission rate based on the scheduling decisions derived in \textbf{P2} and \textbf{P3} is within a constant (arbitrary small if $V$ is large) to the optimum and the virtual queues are mean-rate-stable, where $C$ is a constant related to $A_\textrm{max}$ (\ref{o1}). In the following, we will dig into the methods to solve \textbf{P3}. 
	\end{remark}
	\section{Block Coordinate Update based Method for P3}
	\label{sec_bcu}
	This section is dedicated to solving the scheduling problem of \textbf{P3} only based on the knowledge of statistical CSI. The previous section establishes the optimality of the proposed beam-based scheduling algorithm given the solutions of \textbf{P2} (generally easy to solve) and \textbf{P3}. However, due to the NP-hardness of \textbf{P3}, explicit solutions are hard to attain. More importantly, it is proposed that the scheduling decisions of \textbf{P3} should only rely on statistical CSI, rendering the solution even more intractable. Towards this end, an algorithm based on solving the convex relaxation of the original problem leveraging the BCU technique and random matrix theory is proposed. First, \textbf{P3} is transformed for better exposition based on the uplink-downlink duality \cite{Weingarten06} \cite{Wei04}. The instantaneous achievable rate in \textbf{P3} is evaluated by the MIMO broadcast channel capacity. The following Proposition \ref{prop1} derives an implicit asymptotic expression of the objective function in \textbf{P3} such that the optimization is only dependent on statistical CSI which in this case is the channel correlation matrices.
	\begin{prop}
		\label{prop1}
		In the large system regime, i.e., $K \to \infty$ and $K/N_\textrm{s} \to \beta$, the queue-weighted downlink channel capacity in \textbf{P3} is asymptotically equivalent to 
		\begin{IEEEeqnarray}{rCl}
			\label{conv_relex}
			&& \mathcal{D}(\bm{Q},\bm{R}_1,...,\bm{R}_{N_\textrm{t}},\bm{\Sigma}_\textrm{u},\bm{\Sigma}_\textrm{b},\bm{p}) \nonumber\\
			&\overset{\triangle}{=}& \sum_{n=1}^{N_\textrm{t}} q_{n}  T \log \left(1+ p_{n}s_{n} \textrm{tr} \left[\bm{\Sigma}_\textrm{b} \bm{B}_\textrm{DFT} ^\dag \bm{R}_n \bm{B}_\textrm{DFT} \bm{\Sigma}_\textrm{b} \left(\frac{1}{M}\sum_{j=1}^{n-1} \frac{p_j s_j \bm{\Sigma}_\textrm{b} \bm{B}_\textrm{DFT} ^\dag \bm{R}_j \bm{B}_\textrm{DFT} \bm{\Sigma}_\textrm{b}}{1+e_{n,j}} + \bm{I}\right)^{-1}  \right] \right), \nonumber\\
		\end{IEEEeqnarray}
		where $q_i$'s are arranged in non-increasing order, and $e_{n,i}$ is the unique solution of the following equations.
		\begin{equation}
		\label{ei}
		e_{n,i} = \textrm{tr} \left[\bm{\Sigma}_\textrm{b} \bm{B}_\textrm{DFT} ^\dag \bm{R}_i \bm{B}_\textrm{DFT} \bm{\Sigma}_\textrm{b} \left(\frac{1}{M}\sum_{j=1}^{n-1} \frac{p_j s_j \bm{\Sigma}_\textrm{b} \bm{B}_\textrm{DFT} ^\dag \bm{R}_j \bm{B}_\textrm{DFT} \bm{\Sigma}_\textrm{b}}{1+e_{n,j}} + \bm{I}\right)^{-1}  \right].
		\end{equation}
	\end{prop}
	\begin{IEEEproof}
		See Appendix \ref{app2}.
	\end{IEEEproof}
	\subsection{Convex Relaxation}
	Although the original MIP \textbf{P3} is NP-hard, it can be relaxed to a multi-convex problem by replacing the binary constraints with real-value constraints. The convex relaxation of an MIP is a widely-used technique to achieve near-optimal solutions to the original problem \cite{nai10} \cite{dua06}. The relaxed version of \textbf{P3} is stated below.
	\begin{flalign}
		\label{P4}
		\textbf{P4:}&&\mathop{\textrm{maximize}}\limits_{\bm{\Sigma}_\textrm{b},\bm{w}}  \,\,&  \mathcal{D}(\bm{Q},\bm{R}_1,...,\bm{R}_{N_\textrm{t}},\bm{I}_{N_\textrm{t}},\bm{\Sigma}_\textrm{b},\bm{w}) &&\\
		&&\textrm{s.t.,}\,\,  & \sum_{n=1}^{N_\textrm{t}} w_n \le P,\, \sum_{i=1}^M b_i = K,\,0 \le b_i \le 1, \forall i,&&
	\end{flalign}
	where $w_n = p_n s_n$, and $\mathcal{D}$ is defined in \eqref{conv_relex}. Define the optimum solution of \textbf{P4} as $b_i^*$'s and $w_n^*$'s, respectively. Then the scheduling decision is to schedule the beams and users corresponding to the largest $K$ $b_i^*$'s and $N_\textrm{s}$ $w_n^*$'s, respectively. It is observed that \textbf{P4} is a multi-convex problem \cite{warga63} since the objective function is concave in both $\bm{\Sigma}_\textrm{b}$ and $\bm{w}$. In view of this, the following Algorithm \ref{alg:bcu}, which bases upon the BCU technique is proposed.
	\begin{algorithm}[h]
		\caption{BCU-Based Scheduling}
		\label{alg:bcu}		
		\textbf{Initialization}: $\bm{\Sigma}_\textrm{b}^{(0)} = \bm{I}_M$; \\
		\textbf{Iteration}: \For{ $t=1:T$} {
			\textbf{User scheduling update based on iterative water filling}: $\forall n \in [1,N_\textrm{t}]$, $\omega_n^{(0)} = P/N_\textrm{t}$;
			\For{ $t_{\rm{w}}=1:T_{\rm{w}}$} {
				Compute for each $n$, $\beta_n^{(t_\textrm{w})} = \textrm{tr} \left[\bm{\Sigma}_\textrm{b}^{(t-1)} \bm{B}_\textrm{DFT} ^\dag \bm{R}_n \bm{B}_\textrm{DFT} \bm{\Sigma}_\textrm{b}^{(t-1)} \left(\frac{1}{M}\sum_{j=1}^{n-1} \frac{w_j^{(t_\textrm{w}-1)} \bm{R}_j}{1+e_{n,j}} + \bm{I}\right)^{-1}  \right]$, where $e_{n,i}$ is the unique solution of the equations in \eqref{ei};\\
				Apply the classical water filling algorithm with water levels defined by $\bm{\beta}^{(t_\textrm{w})}$
				\begin{equation}
				\bm{\gamma}^{(t_{\rm{w}})} = \argmax_{\sum_{n=1}^{N_{\rm{t}}} \gamma_n \le P, \bm{\gamma} \ge \bm{0}} \sum_{n=1}^{N_\textrm{t}} q_{n}  \log \left(1+\gamma_n \beta_n^{(t_\textrm{w})}\right) \nonumber
				\end{equation}
				Update $\bm{\omega}$ as $\bm{\omega}^{(t_\textrm{w})} =  \left(1-1/M\right)\bm{\omega}^{(t_\textrm{w}-1)} + (1/M)\bm{\gamma}^{(t_\textrm{w})}$\\
				\If{$\|\bm{\omega}_n^{(t_{\rm{w}})} - \bm{\omega}_n^{(t_{\rm{w}}-1)} \| < \epsilon$}{$\bm{w}^{(t)} = \bm{\omega}^{(t_{\rm{w}})}$, break;}
			}
			\textbf{Beam selection}: Solve for $\bm{\Sigma}_\textrm{b}^{(t)}$, which is the solution to the convex optimization problem of \textbf{P4} with $\bm{w} = \bm{w}^{(t)}$. \\
			\textbf{Stopping criterion}: \If{$\|\bm{w}^{(t)} - \bm{w}^{(t-1)}\| < \epsilon_1$ and $\|\bm{\Sigma}_{\rm{b}}^{(t)} - \bm{\Sigma}_{\rm{b}}^{(t-1)}\| < \epsilon_2$}{$\bm{w}_\textrm{opt} = \bm{\omega}^{(t)}$, $\bm{\Sigma}_\textrm{b,opt}= \bm{\Sigma}_\textrm{b}^{(t)}$, break;}
			
		}
		\textbf{Output}: The scheduling user set is the users with the largest $N_\textrm{t}$ values in $\bm{w}_\textrm{opt}$. The selected beams are the ones with the largest $K$ values in the diagonal entities of $\bm{\Sigma}_\textrm{b,opt}$.		
	\end{algorithm}
	
	The basic idea of the proposed BCU-based user and beam joint scheduling is that an iterative method which cyclically optimizes user scheduling and beam selection with the other fixed is guaranteed to converge to the global optimum of \textbf{P4}. In order to accelerate the iteration, an iterative water filling approach which is based on \cite{jindal05} and deals with user scheduling is adopted. 
	
	\emph{Convergence of the proposed algorithm}: The convergence to the global optimum is due to the convergence results of the BCU algorithm \cite{Tseng01}. The details of the proof is given in Appendix \ref{app3}. It is found through simulations that the BCU-based algorithm normally converges after $2$-$3$ iterations. Therefore, the computational complexity and convergence time are acceptable in simulated scenarios.
	\begin{remark}
		The Algorithm \ref{alg:bcu} can solve the joint user scheduling and beam selection problem based on statistical CSI. Therefore, it is applicable before channel estimations. After the system selects users and beams, the instantaneous channel estimations can be implemented and subsequently digital precoding and decoding can follow. This is in line with the multi-layer signal processing concept proposed in, e.g., \cite{heath16} \cite{Adhikary13}, which proposes that the pre-beamforming should be done based on channel statistics to save RF chains, complexity and system overhead.
	\end{remark}
	\section{Incremental Selection based Method for P3}
	\label{sec_igs}
	Although the BCU-based algorithm is guaranteed to converge to the optimum of the relaxed convex optimization problem, it still has high complexity due to the iterative algorithm design. Therefore it may take a long time to converge. In this regard, an algorithm which selects users and beams incrementally with low complexity is proposed. The key to the design of the algorithm is to derive the incremental selection criterion. Thanks to the results in Proposition \ref{prop1}, the structure of the asymptotic rates in \eqref{conv_relex} can be utilized to give such a criterion. Thereby, the incremental selection algorithm is described in Algorithm \ref{alg:is}, which assumes $K=N_\textrm{s}$\footnote{The assumption is justified by arguing that maximum degree-of-freedom is achieved with $K=N_\textrm{s}$.}.
	\begin{algorithm}[h]
		\caption{Incremental Greedy Scheduling (IGS)}
		\label{alg:is}		
		\textbf{Initialization}: $\mathsf{U} = \mathsf{B} = \emptyset$;  $\mathsf{U}_\textrm{all} = [1:N_\textrm{t}]$, $\mathsf{B}_\textrm{all} = [1:M]$ \\
		\textbf{Incremental Selection}: \For{ $t=1:N_\textrm{s}$} {
			Find the user-$n_t$ and the beam-$b_t$ that maximize:
			\begin{IEEEeqnarray}{rCl}
				(n_t,b_t) &=& \argmax_{n \in \mathsf{U}_\textrm{all} \backslash \mathsf{U},b \in \mathsf{B}_\textrm{all} \backslash \mathsf{B} } \nonumber \\
				&& q_{n}  \log \left(1+ \frac{P}{N_\textrm{s}} \textrm{tr} \left[\bm{\Sigma}_b  \bar{\bm{R}}_n  \bm{\Sigma}_b \left(\frac{1}{M}\sum_{j \in \mathsf{U}} \frac{\frac{P}{N_\textrm{s}}  \bm{\Sigma}_b \bar{\bm{R}}_j \bm{\Sigma}_b}{1+e_{n,j}} + \bm{I}\right)^{-1}  \right] \right) \nonumber
			\end{IEEEeqnarray}
			where $\bar{\bm{R}}_j = \bm{B}_\textrm{DFT} ^\dag \bm{R}_j \bm{B}_\textrm{DFT}$, 
			\begin{equation}
			\left( {{\bm{\Sigma}_b}} \right)_{i,i} = \left\{ {\begin{array}{*{20}{l}}
				{1,\;{\textrm{for}}\;i \in \mathsf{B} \cup {\rm{\{ b\} }}}\\
				{0,\;{\textrm{else,}}}
				\end{array}} \right. \nonumber
			\end{equation}
			and 
			\begin{equation}
			\label{eiis}
			e_{n,i} = \textrm{tr} \left[\bm{\Sigma}_b \bar{\bm{R}}_i \bm{\Sigma}_b \left(\frac{1}{M}\sum_{j \in \mathsf{U}} \frac{ \frac{P}{N_\textrm{s}} \bm{\Sigma}_b \bar{\bm{R}}_j  \bm{\Sigma}_b}{1+e_{n,j}} + \bm{I}\right)^{-1}  \right]. \nonumber
			\end{equation}
			\textbf{Update}: $\mathsf{U} \cup \{n_t\} \rightarrow \mathsf{U}$, $\mathsf{B} \cup \{b_t\} \rightarrow \mathsf{B}$			
		}
		\textbf{Output}: The scheduling user set is $\mathsf{U}$, and the selected beam set is $\mathsf{B}$.	
	\end{algorithm}
	\begin{remark}
		Due to the successive interference cancellation (SIC) structure in the broadcast channel queue-weighted capacity expression in Proposition \ref{prop1}, the rates of the users decoded (selected in IGS) first will not be affected by the users decoded (selected in IGS) later. Therefore, the IGS algorithm is viable because the earlier decisions are decoupled from later ones. 
	\end{remark}
	
	\emph{Complexity analysis}: The IGS has a complexity of $\mathcal{O}(N_\textrm{s}N_\textrm{t}M)$, because in each step it involves an exhaustive search over $\mathcal{O}(M N_\textrm{t})$ possible user and beam combination, and there are $N_\textrm{s}$ iterations. Compared with an exhaustive search over all user and beam subsets with complexity of $\binom{N_\textrm{t}}{N_\textrm{s}}$$\binom{M}{K}$, and the BCU-based scheduling which is difficult to quantize the complexity due to the iterative design of the algorithm, the IGS has a relatively very low complexity.
	
	Due to the greedy nature of the IGS algorithm, it may fail to find the optimum sets of users and beams to schedule. Hence, it is better to have a \emph{worst-case performance bound} of the IGS, such that potentially arbitrarily bad solutions are excluded. Fortunately, this is the case for the proposed IGS, due to the submodularity property \cite{nem78} of the problem. 
	\begin{theorem}
		\label{thm2}
		Denote the queue weighted sum rate achieve by the IGS algorithm as $D_\textrm{IGS}$, and the global optimum as $D_\textrm{opt}$, then it is satisfied that
		\begin{equation}
		D_\textrm{IGS} \ge (1-e^{-1})D_\textrm{opt}.
		\end{equation}
	\end{theorem}
	\begin{IEEEproof}
		The proof is based on the submodularity of the queue-weighted sum rate maximization problem in \textbf{P3}. Informally, the submodularity property indicates the problem has diminishing returns, i.e., in this case the sum rate increase by scheduling a user or a beam is larger when scheduled with a smaller user/beam set, i.e.,
		\begin{equation}
		\label{submodular}
		D(\mathsf{U}_1 \cup u) - D(\mathsf{U}_1 ) \ge D(\mathsf{U}_2 \cup u) - D(\mathsf{U}_2 ),
		\end{equation}
		for any $u \in \mathsf{U}_\textrm{all} \backslash (\mathsf{U}_1 \cup \mathsf{U}_2)$ and $\mathsf{U}_1  \subseteq  \mathsf{U}_2 $. This is easily validated since the same user will suffer from more interference with a larger scheduled user set. There are two additional conditions for submodularity, which is that the function should be \emph{nondecreasing} and \emph{nonnegative}. The nondecreasing property, i.e.,
		\begin{equation}
		\label{nondec}
		D(\mathsf{U} \cup u) \ge D(\mathsf{U}),\,\forall u \in \mathsf{U}_\textrm{all} \backslash \mathsf{U},
		\end{equation}
		can be proved by arguing that at least, zero power can be allocated to the user or beam to obtain equal performance without the user or beam since the objective function is a maximization over all user and beam selection schemes. It should be noted that although the nonnegative condition is easily validated, e.g., for the sum rate maximization or max-min rate maximization, it is not met exactly for the PFS sum logarithm rate optimization. However, if we fix $c_n=1$, $\forall n$ in \eqref{pfs_utility}, the objective function is non-negative and thus the submodularity property is upheld. 
		
		Based on the submodularity, it can be proved that the IGS achieves a near-optimum performance. The remaining details of the proof is well-known in the literature and therefore omitted for brevity \cite{nem78,kim16}.
	\end{IEEEproof}
	\begin{remark}
		Although with Theorem \ref{thm2}, it is only proved that the IGS achieves at least about $60$\% throughput of the optimum scheme, the performance in reality is much better than that, as which will be shown by simulations. Existing work which also utilizes the greedy algorithm with submodularity property also agrees with this finding \cite{kim16}. 
	\end{remark}
	
	\section{Simulation Results}
	\label{sec_sr}
	In this section, simulation results are presented. The channel model is as described in Section \ref{sec.cm}, where the number of MPCs for each user is $L_n = 3$, $\forall n$ (including one LoS MPC), unless stated otherwise. The ULA is used in the simulations and the amplitude of the LoS MPC is $10$ times the one of the NLoS MPCs. The DoAs of the signals are generated from i.i.d. uniform distributions. The antenna spacing $d = \lambda/2$, where $\lambda$ denotes the carrier wavelength. In some of the following cases where users' pathlosses are not identical, the distances of users are generated based on an i.i.d. uniform distributions from $30$ to $200$ meters and the pathloss $\gamma_n$ is
	\begin{equation}
	\gamma_n = \left(\frac{d_n}{d_0}\right)^{-\gamma},
	\end{equation}
	where $\gamma=2$ which is in line with mm-wave channel measurements \cite{rap13} and $d_0$ is some reference point distance. 
	The regularized zero-forcing (RZF) precoder is adopted for system evaluation, i.e., define $\bm{K}_{\textrm{rzf}} = \left(\bm{\bar H}^\dag\bm{ {\bar H}}+M\alpha \bm{I}_M\right)^{-1}$. The RZF precoding matrix is expressed as $\bm{B}_{\textrm{d}} = \zeta \bm{K}_{\textrm{rzf}} \bm{ {\bar H}}^\dag$, where $\zeta$ is a normalization scalar to fulfill the power constraint, and $\alpha$ is the regularization factor \cite{peel05}. Although RZF precoder is not the optimal coding scheme for Gaussian broadcast channel (dirty-paper-coding with minimum-mean-square-error (MMSE) precoder is proved for optimality but limited in reality due to high complexity), it can achieve full degree-of-freedom (DoF) in the high signal-to-noise-ratio (SNR) region and is easy to implement. In the simulations, $\alpha = N_\textrm{s}/\rho$, where $\rho$ is the receive SNR \cite{peel05}. The user instantaneous rate is calculated by the Shannon formula. The block fading model is adopted, where the channel stays constant for $10$ time slots and evolves to another realization based on an i.i.d. distribution. The phase and amplitude of each MPC is generated randomly. The simulation runs for $1000$ such blocks and calculate the time-averaged downlink transmission rates. The constants used in the Lyapunov-drift optimization are set to be $V=A_\textrm{max}=10^2 r_\textrm{max}$, where $r_\textrm{max}$ is the maximum rate of the users. The $\epsilon$, $\epsilon_1$ and $\epsilon_2$ in the stopping criterion in BCU-based algorithm are set to be $10^{-2}\rho/K$. In comparisons, the state-of-the-art interference aware beam selection scheme (IA-BS) in \cite{gao16_bs}, and the BDMA scheme in \cite{sun15} are also simulated. These schemes for comparisons are shown to perform better than several other existing schemes, e.g., \cite{ama15}.
	
	\begin{figure}[!t]
		\centering
		\includegraphics[width=0.5\textwidth]{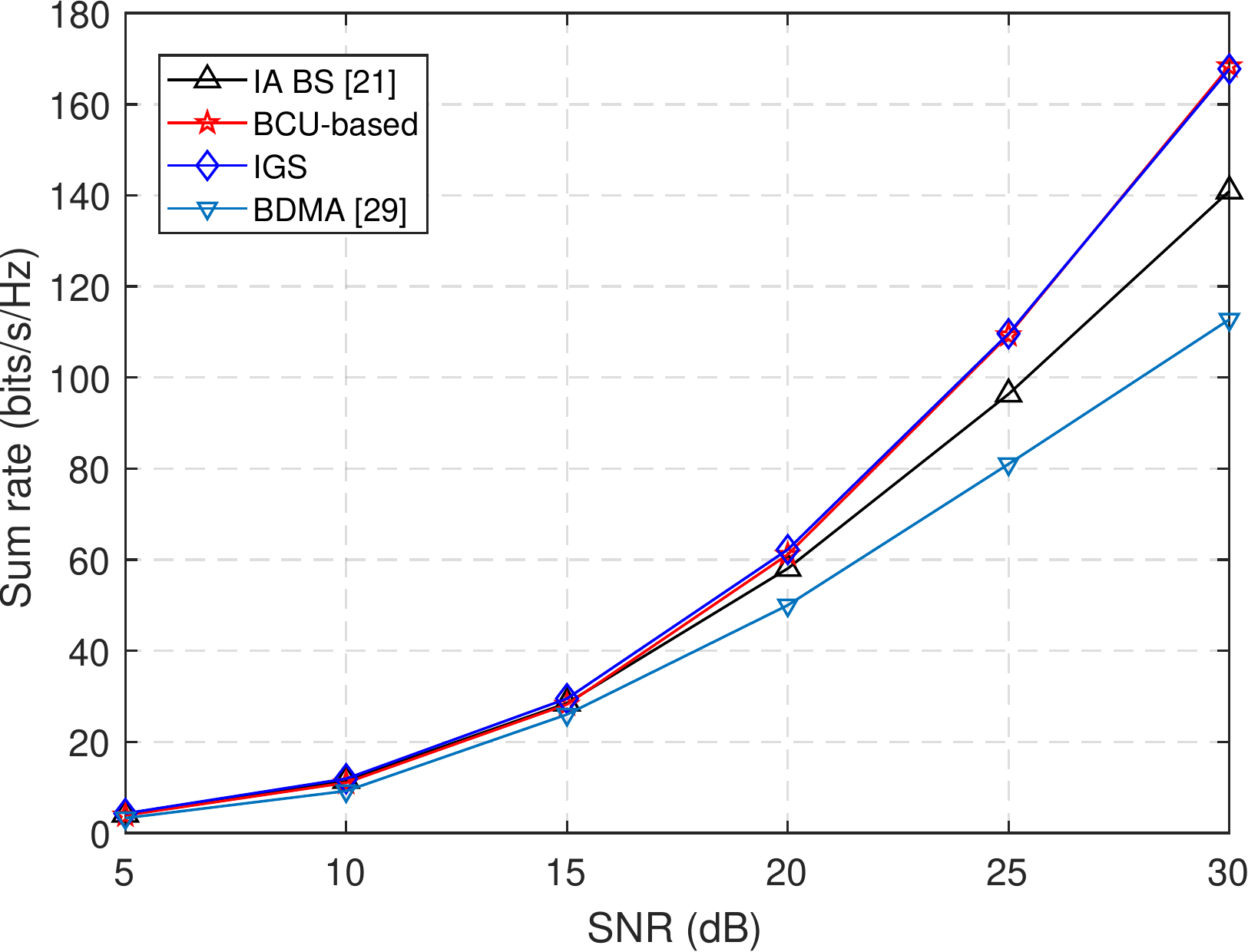}
		\caption{Sum rate comparisons with identical user pathloss. The number of BS antennas is $256$, the number of users is $100$, the number of scheduled users and beams are both $40$.}
		\label{Fig_sum_rate_snr_msr_nopathloss}
	\end{figure}
	In Fig. \ref{Fig_sum_rate_snr_msr_nopathloss}, our proposed BCU-based algorithm and the IGS algorithm are compared with the IA-BS algorithm \cite{gao16_bs} and the BDMA scheme \cite{sun15}. Since the sum rates are considered, the utility function in \eqref{sr_utility} is adopted. The scheduled user set in IA-BS is assumed to have the most channel power. It is observed that by jointly considering user scheduling and beam selection, the sum rate performance can be improved in the high-SNR regime. The reason is that in the high-SNR regime, the interference is dominating the performance, and thus by jointly considering the user scheduling and beam selection by the proposed schemes, the interference is better suppressed. The BDMA scheme simply adopts a sequential approach which selects the beam-user pairs incrementally, and it builds on optimizing a sum-rate upper bound; both factors lead to performance degradation. Thus, the resultant performance is not as good. Nonetheless, it should be noted that the BDMA scheme is designed for multiple-antenna users and hence the interference can be suppressed thereby whereas such effects are not captured in the presented simulations. Therefore, we focus on the IA-BS scheme for comparisons in the following. Furthermore, the figure also shows that the BCU-based and IGS algorithms achieve very similar performance in this scenario.
	
	\begin{figure}[!t]
		\centering
		\begin{minipage}[c]{0.5\textwidth}
			\includegraphics[width=8cm]{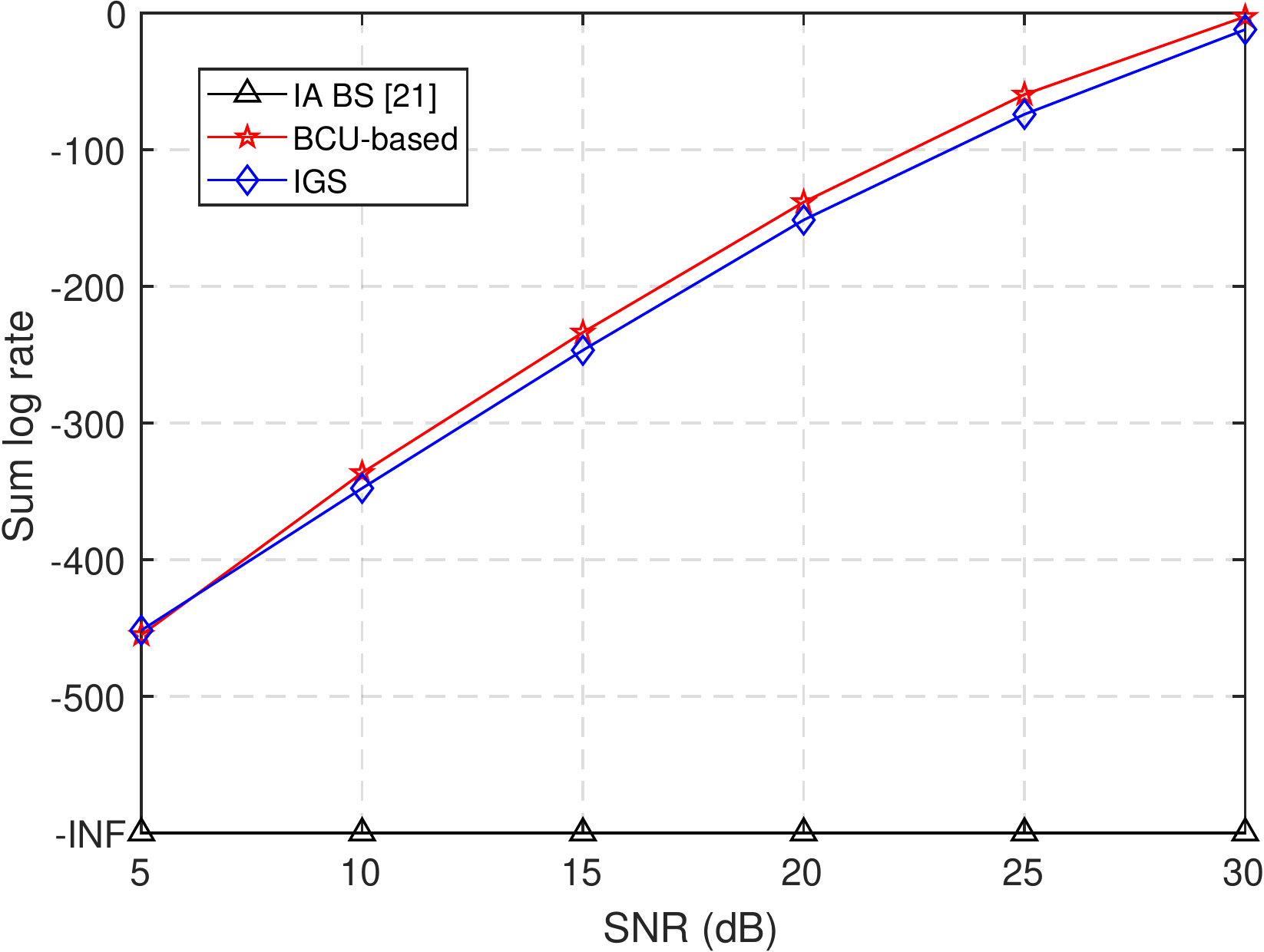}	
		\end{minipage}%
		\begin{minipage}[c]{0.5\textwidth}
			\centering
			\includegraphics[width=8cm]{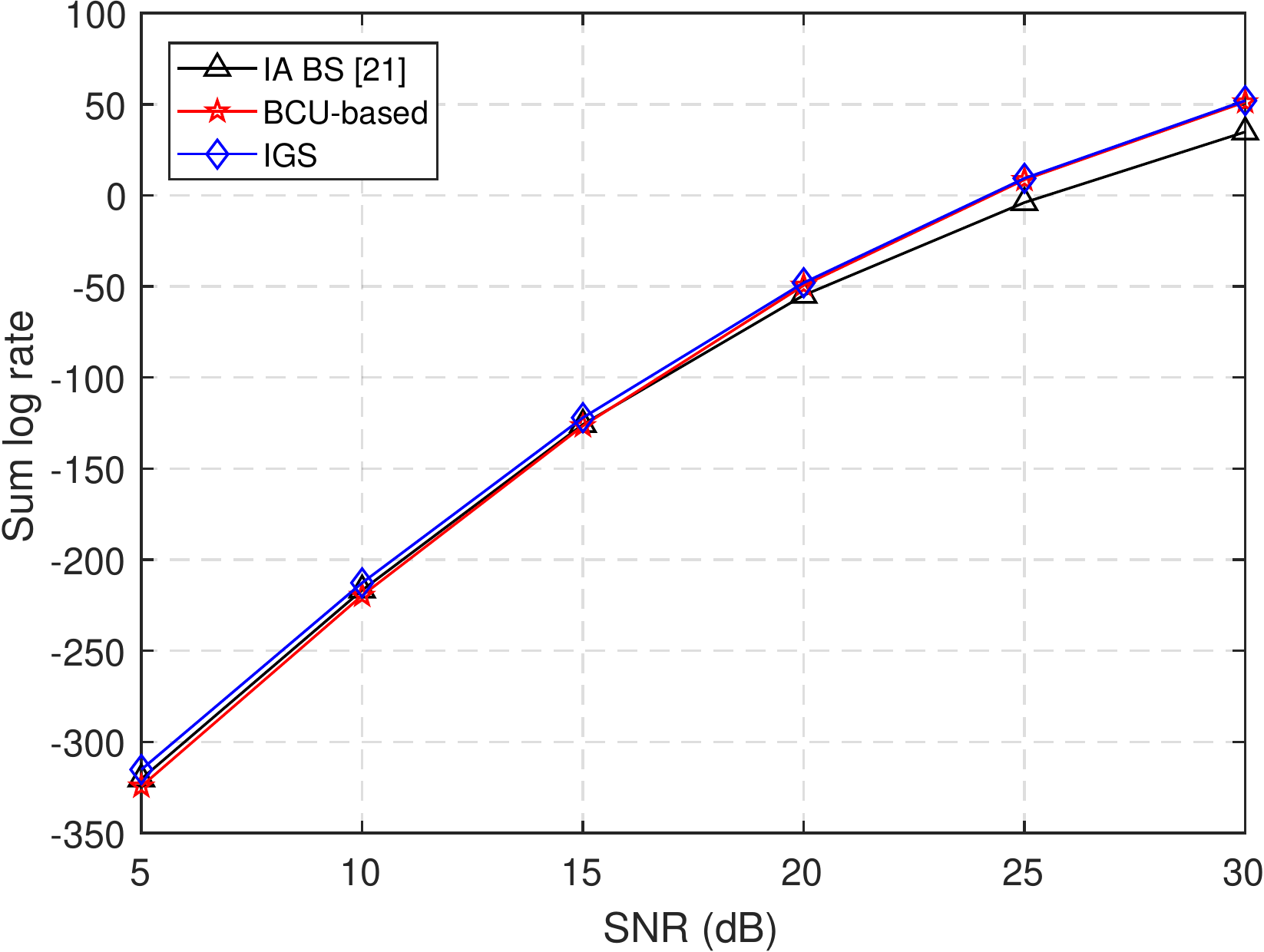}
		\end{minipage}
		\caption{Sum log rate comparisons with (left) and without (right) non-identical user pathloss. The users with non-identical pathloss are generated with distances i.i.d. uniformly distributed between $20$~m to $200$~m. The number of BS antennas is $256$, the number of users is $100$, the number of scheduled users and beams are both $40$.}
		\label{Fig_sum_log_rate_snr_pfs_pathloss}
	\end{figure}
	Considering the utility functions with user-fairness considerations, e.g., the PFS utility function in \eqref{pfs_utility}, the performance advantage of the proposed schemes is more obvious as shown in Fig. \ref{Fig_sum_log_rate_snr_pfs_pathloss} (left). Based on the IA-BS, the beams with stronger channels are always preferred, corresponding to users with small pathloss, resulting in ignorance of the fairness among users. In the proposed joint scheduling schemes, the admission control in \textbf{P2} utilizes a virtual queue to control the fairness among users. Note that since there are always unscheduled users in the IA-BS due to their small pathloss, the sum log rate of the IA-BS scheme in this case is negative infinity. In Fig. \ref{Fig_sum_log_rate_snr_pfs_pathloss} (right), the performance with identical pathloss is presented. Due to the fact that users have identical large-scale fading, and hence equal probability to have good channels, the IA-BS can achieve reasonably good fairness performance.
	
\begin{figure}[!t]
		\centering
		\begin{minipage}[c]{0.5\textwidth}
			\centering
			\label{Fig_sum_rate_snr_msr_nopathloss_opt} 
			\includegraphics[width=8cm]{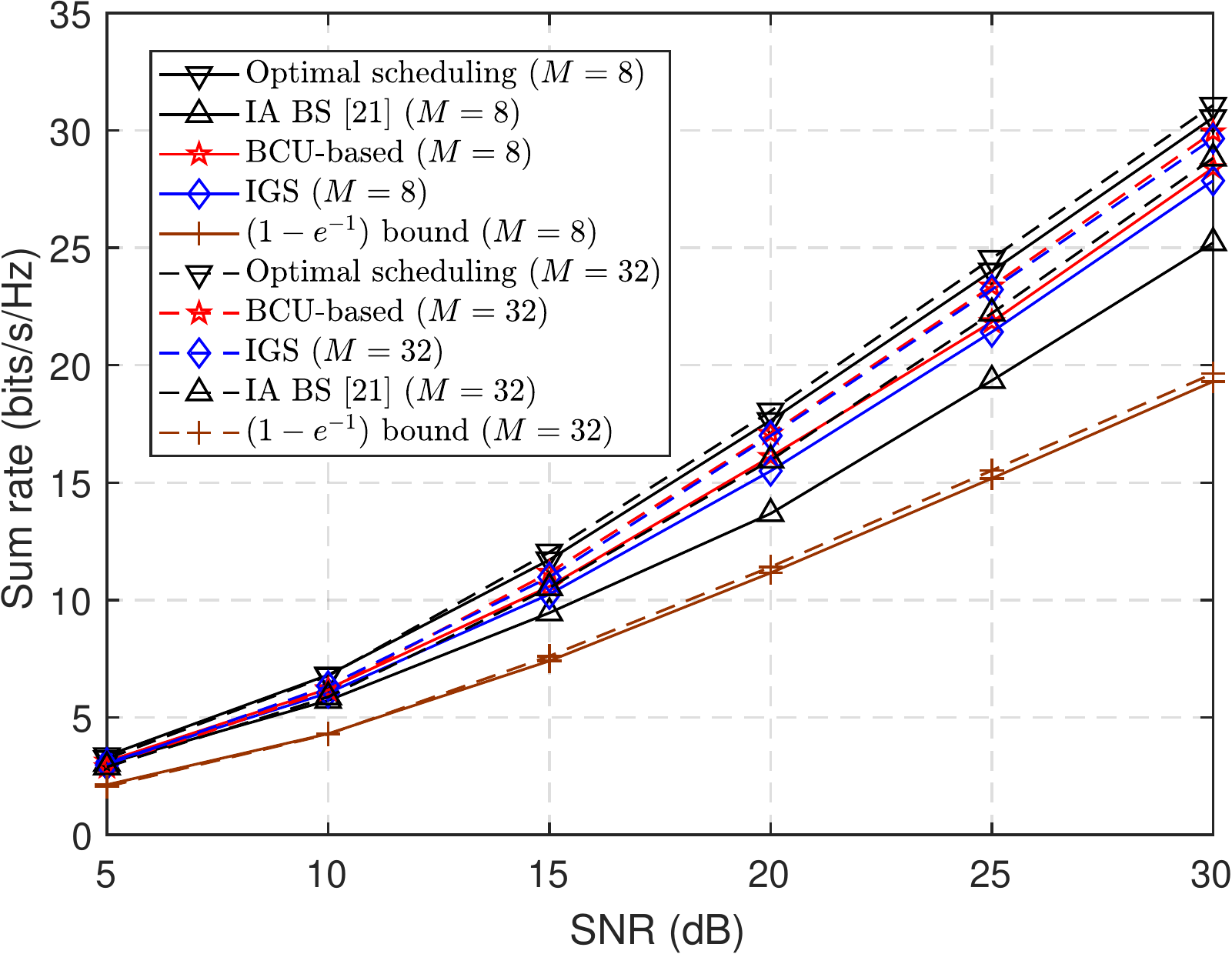}
		\end{minipage}%
		\begin{minipage}[c]{0.5\textwidth}
			\centering
			\label{Fig_sum_log_rate_snr_pfs_pathloss_opt} 
			\includegraphics[width=8cm]{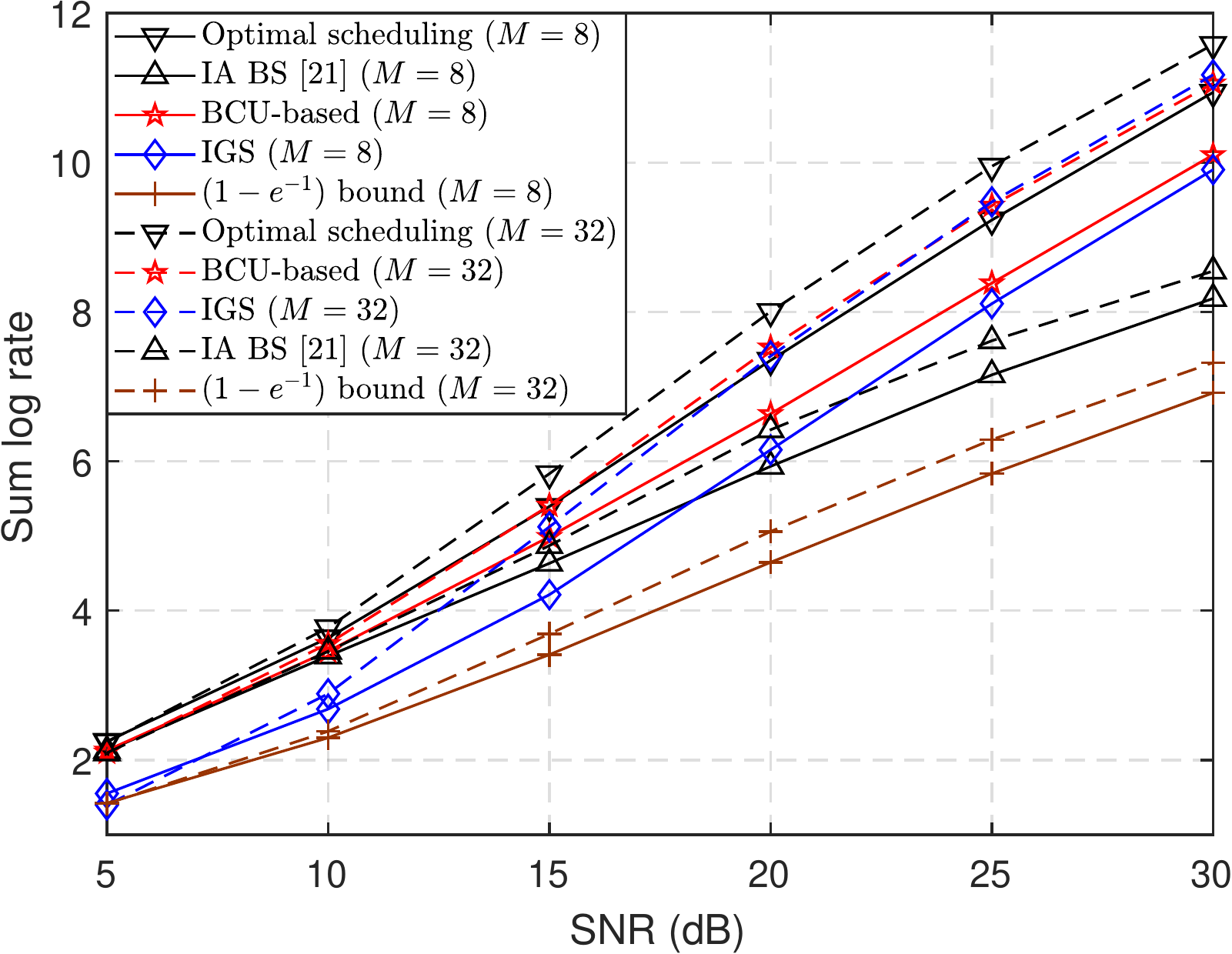}
		\end{minipage}
		\caption{Comparisons with optimal scheduling. The number of users is $8$, the number of scheduled users and beams are both $4$. $c_n=1$, $\forall n$ as in \eqref{pfs_utility}.}
		\label{Fig_opt}
	\end{figure}

	In order to show the performance loss of our proposed schemes compared with optimal scheduling, an exhaustive search over all the feasible user and beam sets is conducted to solve \textbf{P3} and the optimal scheduling performance is obtained accordingly. Due to the prohibitive high complexity of exhaustive search, we consider a small-scale problem where there are $8$ users, $8$ BS antennas and $4$ scheduled users and beams. Nonetheless, it is found by many existing works, e.g., \cite{hoch04} \cite{4}, that the impact of imperfect downlink scheduling decreases with the increase of antenna dimension due to the channel hardening effect. Therefore, the relative performance gap with a larger system dimension should be smaller, or at least similar with that in Fig. \ref{Fig_opt}. In Fig. \ref{Fig_opt}, the left and right figures show sum-rate and sum-log-rate optimizations, respectively. It is observed that in general the proposed schemes can achieve near-optimal performance. The BCU-based scheme is shown to have better performance compared with the IGS, but with higher complexity. The performance bound which we prove for the IGS algorithm is also plotted in the figure. The IGS scheme always performs better than the $(1-1/e)$ bound. In the low SNR regime, it is observed that the IA-BS scheme outperforms the IGS scheme, due to the reason that the IA-BS scheme always selects the user and its corresponding beam which have the strongest channel, and that in Fig. 4 we set $c_n=1$ in the log rate to ensure positive utilities and thus less penalty on the unfairness among users is accounted for.
	
\begin{figure}[!t]
		\centering
		\begin{minipage}[c]{0.5\textwidth}
			\centering
			\label{Fig_l1} 
			\includegraphics[width = 8cm]{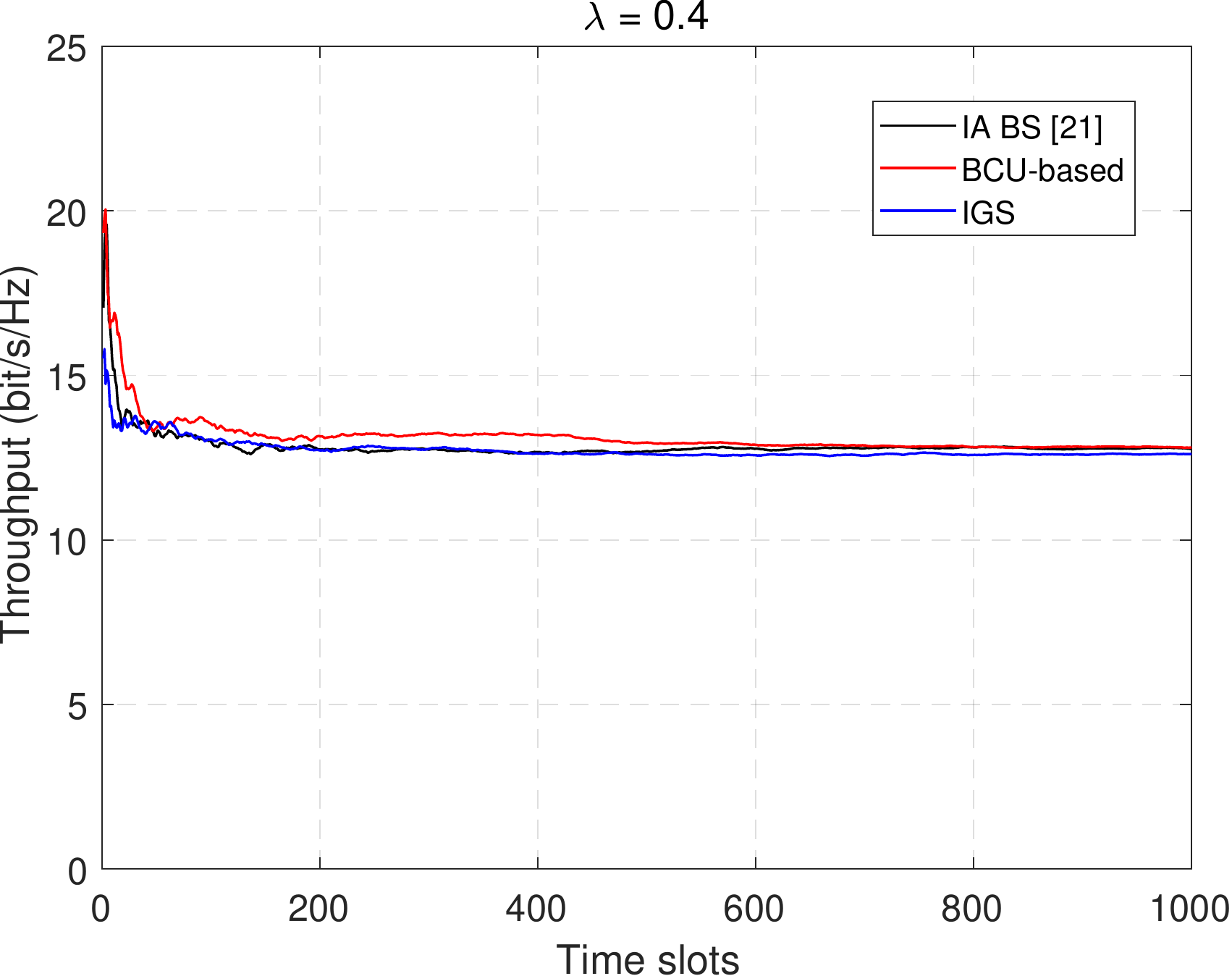}
		\end{minipage}%
		\begin{minipage}[c]{0.5\textwidth}
			\centering
			\label{Fig_l2} 
			\includegraphics[width = 8cm]{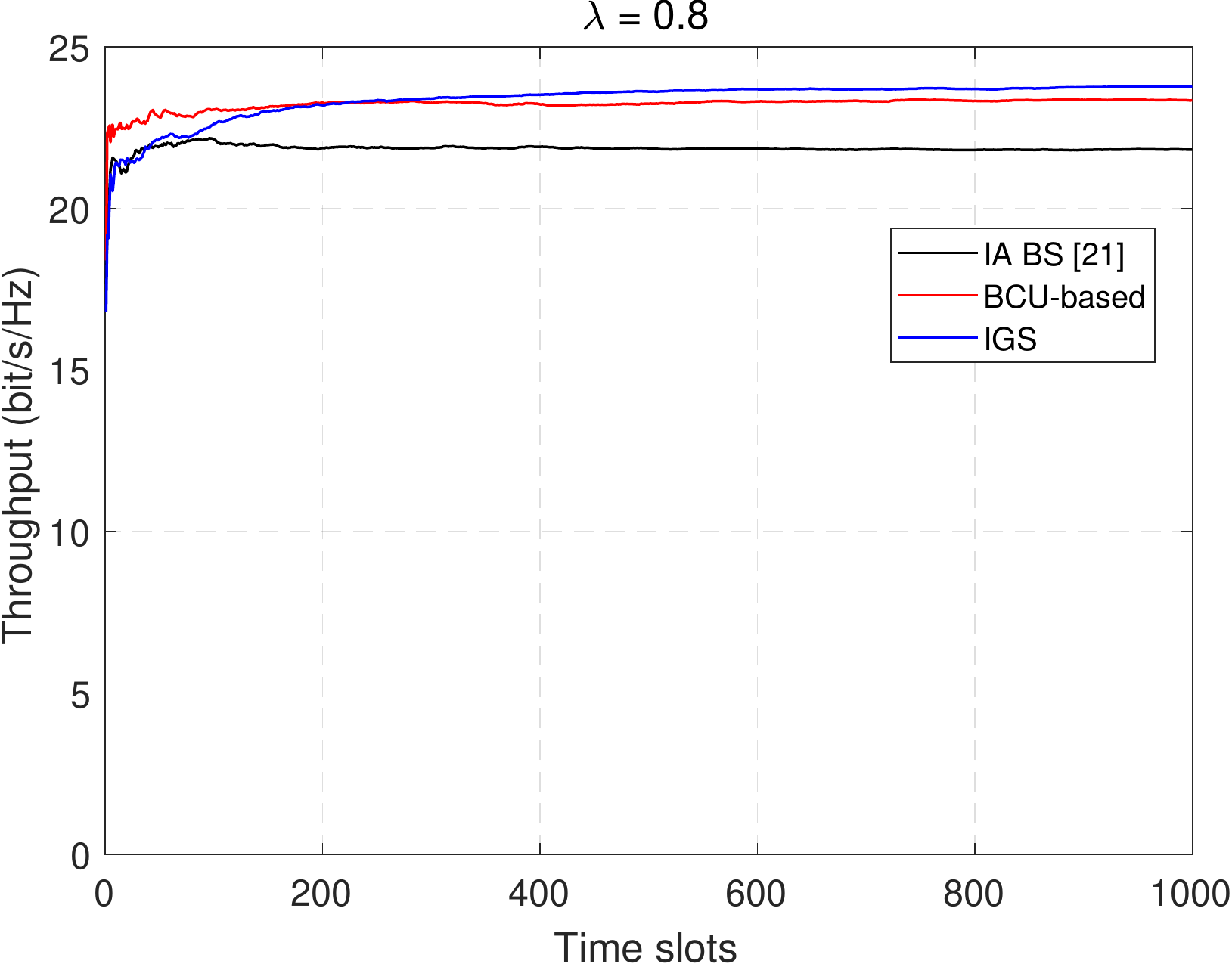}
		\end{minipage}
		\caption{Sample paths of the system average throughput evaluations. The number of BS antennas is $64$, the number of users is $40$, the number of scheduled users and beams are both $20$. $\eta=0.4$.}
		\label{Fig_traffic}
	\end{figure}

\begin{figure}[!t]
		\centering
		\includegraphics[width=0.5\textwidth]{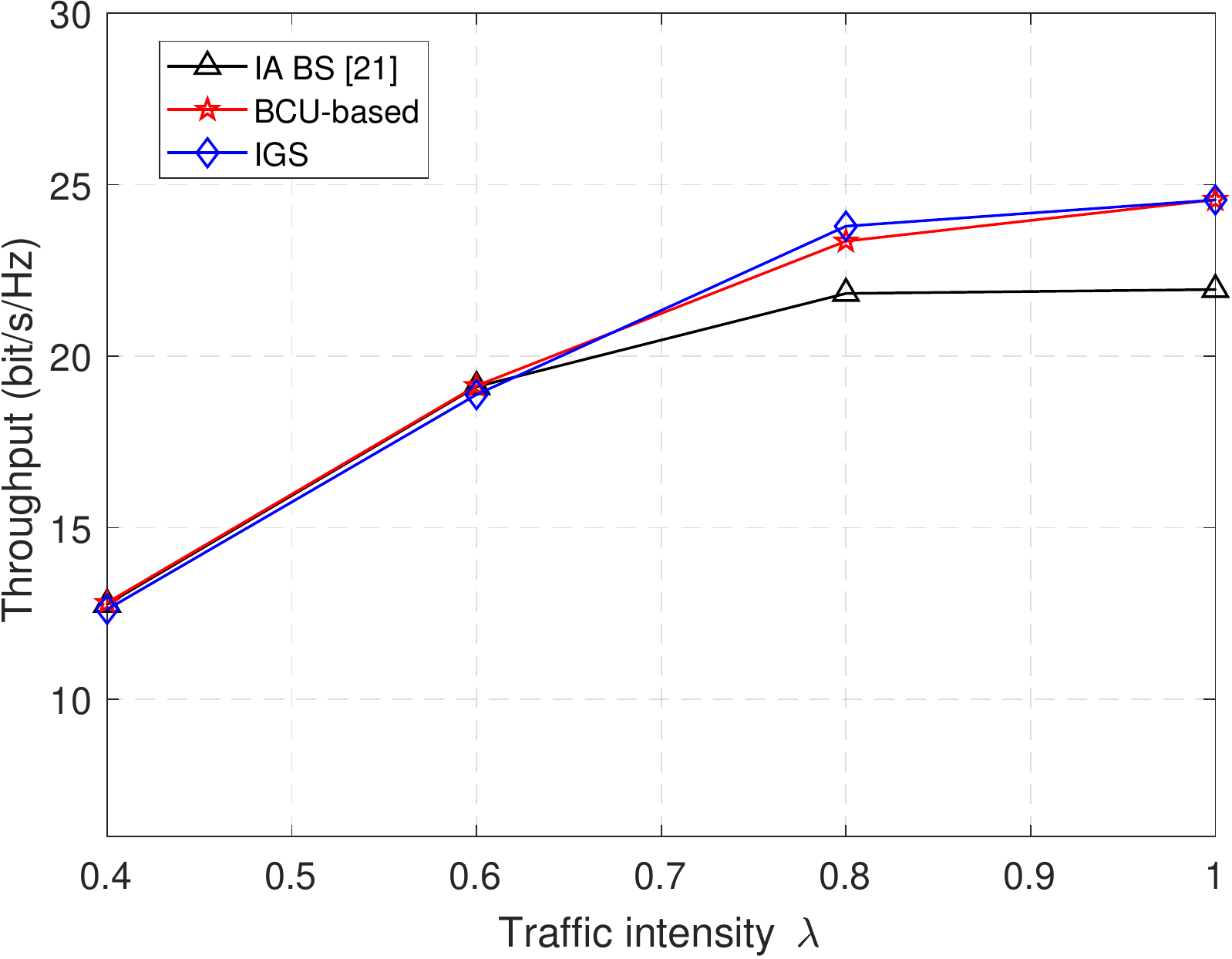}
		\caption{The number of BS antennas is $64$, the number of users is $40$, the number of scheduled users and beams are both $20$. $\eta=0.4$.}
		\label{Fig_lambda}
	\end{figure}

	A throughput comparison with a stochastic traffic model is carried out and shown in Fig. \ref{Fig_traffic} and \ref{Fig_lambda}. Instead of considering full-buffer greedy transmitting sources, each user's arrival traffic is modeled as a Bernoulli process, i.e., the arrival process for user-$n$ 
	\begin{equation}
	\alpha_n = b_n r_\textrm{c},\,\forall n
	\end{equation}
	where $b_n$ is i.i.d. Bernoulli distributed with expected mean values of $p_n$, $r_\textrm{c}$ is a constant which denotes approximately the service rate of each user, and $r_\textrm{c} = \frac{N_\textrm{s}}{N_\textrm{t}}\log(1+\eta \frac{\rho}{ N_\textrm{s}})$ where $\eta$ denotes the approximate SNR loss coefficient introduced by interference. Therefore, based on this setting, the mean values of $b_n$ can be regarded as the traffic intensity where $p_n=0$ denotes zero traffic, and $p_n$ close to $1$ denotes heavy traffic. The queuing dynamic is, with slightly abusing the notations,
	\begin{equation}
	\label{Queueactual}
	{q_n}(t+1) = {q_n}(t) - {\tilde \mu _n}(t) + {\alpha_n}(t),
	\end{equation}
	where  ${{\tilde \mu }_n}(t) = \min \{ {q_n}(t),{\mu _n}(t)\}$ denotes the actual service rate taken into account of empty queues. The throughput is calculated by averaging the sum actual service rate of each user ${{\tilde \mu }_n}(t)$. 
	
	Our proposed Lyapunov-drift based schemes, i.e., BCU-based and IGS, can be easily adapted to this traffic model, by replacing the virtual queues in \eqref{QueueD} with \eqref{Queueactual} and eliminating the admission control step in \textbf{P2}. We assume the IA-BS scheme \cite{gao16_bs} schedules $N_\textrm{s}$ users with the $N_\textrm{s}$-largest queue lengths and selects beams accordingly, which is in line with the methodology that upper layers, e.g., medium-access-control (MAC) layer, schedules some users and push the bits to physical layer. In comparisons, the proposed schemes jointly considers user scheduling with traffic demands and beam selection. One sample path of the system is depicted in Fig. \ref{Fig_traffic}, where the left and right sub-figures denote relatively low and heavy traffic, respectively. The system average throughput is stationary after about $100$ time slots, which equals about $100$~ms given the LTE numerologies where one time slot (transmission-time-interval) is one subframe ($1$~ms). With this convergence time, it is found that the proposed schemes can effectively converge to a reasonably good solution before the statistical CSI varies (usually at a time scale of several seconds).
	In Fig. \ref{Fig_lambda}, the average throughput is compared among different scheduling schemes under different traffic intensities. It is observed that the proposed schemes outperform the IA-BS scheme when the system is with high traffic load. Note that when the system is not fully loaded, i.e., traffic intensity is lower than about $0.6$, the average throughput equals the sum arrival rate and thus the performance advantage of the proposed schemes have not emerged. It is shown that joint considerations of user scheduling and beam selection leads to better system throughput.
	
	\section{Conclusions}
	\label{sec_cl}
	In this paper, the BCU-based scheduling scheme and the IGS scheme are proposed to address the join user scheduling and beam selection optimization problem in beam-based massive MIMO systems based only on statistical CSI. The problem is formulated under the Lyapunov-drift optimization framework. In order to solve the weighted rate maximization problem therein, the proposed BCU-based scheme leverages the convex relaxation of the problem and adopts the BCU technique with the iterative water-filling approach. It is proved that the BCU-based scheduling scheme iteratively converges to the optimum of the relaxed problem. Due to its iterative algorithm structure, relatively high complexity is required. Towards this end, the IGS algorithm is proposed which is based on a greedy approach. Nevertheless, it is proved that the IGS scheme can achieve performance within a multiplicative factor of $(1-e^{-1})$ to the optimum. In simulations, it is shown that the proposed schemes can achieve near-optimal performance and outperform the state-of-the-art beam selection schemes, with utilities such as sum rate and proportional fairness. While existing works focuses on the beam selection, which effectively strive to maximize the sum rate performance, they are not optimized when the user scheduling and beam selection are jointly considered especially when user fairness is taken into account. The performance bound we derive for the IGS scheme is also shown to be well observed.
	\appendices
	\section{Proof of Theorem \ref{thm1}}
	\label{app_1}
	We first briefly review the Lyapunov-drift approach, which is the main mathematical tool in our proof. Define the Lyapunov function as
	\begin{equation}
	\label{Lfunction}
	L(t) \triangleq \frac{1}{2}\sum\limits_n {Q_n^2({t})},
	\end{equation}
	and the Lyapunov drift as
	\begin{equation}
	\label{Ldrift}
	\Delta (t)\triangleq \mathbb{E}[L(t+1)-L(t)|\bm{Q}(t)].
	\end{equation}
	\begin{lemma}
		\label{lemma1}
		If there exist constants $B$ and $\epsilon$, which satisfy
		\begin{equation}
		\label{dstability}
		\Delta (t) \le B -\epsilon \sum\limits_n {Q_n({t})},
		\end{equation}
		then we have:
		
		1) If $\epsilon \ge 0$, then all queues are mean rate stable.
		
		2) If $\epsilon > 0$, then
		\begin{equation}
		\label{QueueStk}
		\mathop {\lim \sup }\limits_{\tau \to \infty } \frac{1}{\tau} \sum\limits_{t = 1}^\tau  \mathbb{E}\left[\sum_n {{Q_n}({t})} \right] \le \frac{B}{\epsilon},
		\end{equation}
		and hence all queues are strongly stable.
	\end{lemma}
	\begin{IEEEproof}
		The proof of Lemma \ref{lemma1} follows the standard procedure as in \cite{neely10}.
	\end{IEEEproof}
	Given the queuing dynamics \eqref{QueueD} and based on the definition in \eqref{Ldrift}, we have
	\begin{IEEEeqnarray}{rCl}
		\label{thmp1}
		\Delta ({t}) &\le& \mathbb{E}\left[\frac{1}{2}\sum_n \left.\left(Q^2_n(t+1) - Q^2_n(t)\right) \right|\bm{Q}(t)\right] \nonumber\\
		&=& \mathbb{E}\left[\frac{1}{2}\sum_n \left.\left(\left({Q_n}(t) - {\tilde \mu _n}(t) + {a_n}(t)\right)^2 - Q^2_n(t)\right) \right|\bm{Q}(t)\right] \nonumber\\
		&=& \mathbb{E}\left[\sum_n \left.\left(\frac{1}{2}{\tilde \mu _n^2}(t) - {\tilde \mu _n}(t){Q_n}(t) + \frac{1}{2}{a_n^2}(t) + {Q_n}(t) {a_n}(t) - {a_n}(t) {\tilde \mu _n}(t) \right) \right|\bm{Q}(t)\right] \nonumber\\
		&\le& \mathbb{E}\left[\sum_n \left.\left(\frac{1}{2} { \mu _n^2}(t) - {\mu _n}(t){Q_n}(t) + \frac{1}{2}{a_n^2}(t) + {Q_n}(t) {a_n}(t) \right) \right|\bm{Q}(t)\right] \nonumber\\
		&=& \mathbb{E}\left[\sum\limits_n \left. {\frac{{\mu _n^2({t}) + a_n^2({t})}}{2}} \right|\bm{Q}({t})\right]  - \sum\limits_n {{Q_n}({t})} \mathbb{E}\left[{\mu _n}({t}) - {a_n}({t})|\bm{Q}({t})\right].
	\end{IEEEeqnarray}
	Observing that
	\begin{IEEEeqnarray}{rCl}
		\label{thmp2}
		\mathbb{E}&&\left[\sum\limits_n \left. {\frac{{\mu _n^2({t}) + a_n^2({t})}}{2}} \right| \bm{Q}({t})\right] \le \frac{T^2}{2}\left[r_{n,\textrm{max}}+ A_{\textrm{max} }^2 \right] \triangleq B,
	\end{IEEEeqnarray}
	where $r_{n,\textrm{max}} = \log\left(1+\|\bm{h}_n\|^2 P\right)$ denotes the maximum transmission rate in one channel use since $r_{n,\textrm{max}}$ is the channel capacity as if the user $n$ was alone, it follows that
	\begin{IEEEeqnarray}{rCl}
		\label{d1}
		\Delta ({t}) &\le& B  - \sum\limits_n {{Q_n}({t})} \mathbb{E}\left[{\mu _n}({t}) - {a_n}({t})|\bm{Q}({t})\right].
	\end{IEEEeqnarray}
	Therefore, for any arrival rate inside the ergodic capacity region, since the scheduling problem in \textbf{P3} minimize the right-hand side of \eqref{d1}, the condition in Lemma \ref{lemma1} \eqref{dstability} is upheld with $\epsilon \ge 0$, i.e., all queues are mean rate stable. In order to show the throughput-optimality in \eqref{opt}, subtract a term related to the utility function,
	\begin{IEEEeqnarray}{rCl}
		\label{e1}
		\Delta ({t}) -V \mathbb{E} \left[\left.U(\bm{a}(t)) \right| \bm{Q}(t)\right] &\le& B - \underbrace{ \sum\limits_n {{Q_n}({t})} \mathbb{E}\left[{\mu _n}(t) | \bm{Q}(t)\right] }_{\textbf{Scheduling}} \nonumber \\
		&+& \underbrace{\mathbb{E}\left[\left.\sum_n {Q_n(t)a_n}({t})-V U(\bm{a}(t)) \right|\bm{Q}({t})\right]}_{\textbf{Admission control}},
	\end{IEEEeqnarray}
	It is observed that the admission control and scheduling problems in \textbf{P2} and \textbf{P3} are equivalent to minimize the related terms in \eqref{e1} as labeled. Therefore, given the solution to \textbf{P2} and \textbf{P3}, the left-hand side of \eqref{e1} is less than the term on the right-hand side with any queue-independent scheduling and admission control. Concretely,
	\begin{IEEEeqnarray}{rCl}
		\label{f1}
		\Delta ({t}) -V \mathbb{E} \left[\left.U(\bm{a}(t)) \right| \bm{Q}(t)\right] &\le& B - \sum\limits_n {{Q_n}({t})} \bar {R}_n+\sum_n {Q_n(t)z_n}-V U(\bm{z}) ,
	\end{IEEEeqnarray}
	where $\bar {R}_n$ and $z_n$ denotes any queue-independent service rate and admission rate, respectively. Taking expectations on both sides over $\bm{Q}(t)$, and taking the telescoping sum yields (assuming $
	\bm{Q}(0)=0$ for better exposition),
	\begin{IEEEeqnarray}{rCl}
		\label{g1}
		\frac{1}{\tau} \sum\limits_{t = 0}^{\tau-1}  \mathbb{E}\left[ {{Q_n}({t})} \right](\bar {R}_n - z_n) &\le& B + V\left(U\left(\frac{1}{\tau} \sum\limits_{t = 0}^{\tau-1}\mathbb{E}\left[\bm{a}(t)\right]\right) - U(\bm{z})\right)
	\end{IEEEeqnarray}
	Let $\bm{z} = \bar{\bm{R}}^*$ which is the rate point in $\mathcal{R}$ that achieves the optimum utility function. Based on the fact that all queues are mean rate stable as shown before, the left-hand side is non-negative, it then follows that 
	\begin{IEEEeqnarray}{rCl}
		\label{o1}
		U(\bar{\bm{R}}^*) &\le& U\left(\frac{1}{\tau} \sum\limits_{t = 0}^{\tau-1} \mathbb{E}\left[\bm{a}(t)\right]\right) + B/V \nonumber\\
		&\le& U\left(\frac{1}{\tau} \sum\limits_{t = 0}^{\tau-1} \mathbb{E}\left[\bm{R}(t)\right]\right) + B/V
	\end{IEEEeqnarray}
	Let $\tau \to \infty$, take the $\limsup$ and rearrange the terms yields the optimality condition in \eqref{opt}. The inequality in \eqref{o1} is based on the fact the queues are all mean rate stable the utility function is non-decreasing. This completes the proof.

	\section{Proof of Proposition \ref{prop1}}
	\label{app2}
	Let $q_n = Q_n(t)$ denote the virtual queue state at the scheduling time $t$, define
	\begin{IEEEeqnarray}{rCl}
		\mathcal{G}(\bm{Q},\bm{H},\bm{\Sigma}_\textrm{u},\bm{\Sigma}_\textrm{b},\bm{p})&\overset{\Delta}{=}& \sum_{n=1}^{N_\textrm{t}} q_n  s_n\sum_{t=1}^T R_n\left(\bm{H}(t),\bm{\Sigma}_\textrm{u},\bm{\Sigma}_\textrm{b},\bm{p}\right) \nonumber \\
		&=& \sum_{n=1}^{N_\textrm{t}} q_{\pi_n}  s_{\pi_n}\sum_{t=1}^T \log\frac{\det\left(\bm{I}+\sum_{j=1}^n \bm{\Sigma}_\textrm{b} \bm{B}_\textrm{DFT}^\dag \bm{h}_{\pi_j}\bm{h}_{\pi_j}^\dag \bm{B}_\textrm{DFT} \bm{\Sigma}_\textrm{b}  p_{\pi_j}s_{\pi_j}\right)}{\det\left(\bm{I}+\sum_{j=1}^{n-1} \bm{\Sigma}_\textrm{b} \bm{B}_\textrm{DFT}^\dag\bm{h}_{\pi_j}\bm{h}_{\pi_j}^\dag \bm{B}_\textrm{DFT} \bm{\Sigma}_\textrm{b} p_{\pi_j}s_{\pi_j}\right)} \nonumber \\
		&=& \sum_{n=1}^{N_\textrm{t}} q_{\pi_n} s_{\pi_n}  \sum_{t=1}^T \log \left(1+ \bm{h}_{\pi_n}^\dag \bm{B}_\textrm{DFT} \bm{\Sigma}_\textrm{b} \bm{A}^{-1}  \bm{\Sigma}_\textrm{b} \bm{B}_\textrm{DFT}^\dag \bm{h}_{\pi_n} p_{\pi_n}s_{\pi_n} \right) \nonumber \\
		&=& \sum_{n=1}^{N_\textrm{t}} q_{n}  \sum_{t=1}^T \log \left(1+ \bm{h}_{n}^\dag \bm{B}_\textrm{DFT} \bm{\Sigma}_\textrm{b} \bm{A}^{-1} \bm{\Sigma}_\textrm{b} \bm{B}_\textrm{DFT} ^\dag \bm{h}_{n} p_{n}s_{n} \right)
	\end{IEEEeqnarray}
	where 
	\begin{IEEEeqnarray}{rCl}
		\bm{A} &=& \bm{I}+ \bar{\bm{H}}_{[n-1]} \bar{\bm{H}}_{[n-1]}^\dag, \nonumber\\
		\bar{\bm{H}}_{[n-1]} &=&  \sqrt{ p_{\pi_j}s_{\pi_j}} \bm{\Sigma}_\textrm{b} \bm{B}_\textrm{DFT}^\dag \left[\bm{h}_{1},...,\bm{h}_{n-1}\right],
	\end{IEEEeqnarray}
	and $\pi_i \in [1,...,{N_\textrm{t}}]$  is a permutation of the user index which satisfies $q_{\pi_1}  s_{\pi_1} \ge ... \ge q_{\pi_{N_\textrm{t}}}  s_{\pi_{N_\textrm{t}}}$ representing the decoding order in the dual uplink multiple-access-channel \cite{Wei06}. The last equality is based on the fact that 
	\begin{equation}
	x\log\det(\bm{I}+\bm{A}x) = \log\det(\bm{I}+\bm{A}x), \textrm{ } \forall x \in \{0,1\},
	\end{equation}
	and without loss of generality, we assume $q_i$'s are arranged in non-increasing order. We invoke \cite[Theorem 1]{Wagner12} which is stated at the end of the proof as Lemma \ref{lm2} for reading convenience. Denote the channel correlation matrix for user-$n$ as $\bm{R}_n$, and it yields
	\begin{IEEEeqnarray}{rCl}
		&& \mathcal{G}(\bm{Q},\bm{H},\bm{\Sigma}_\textrm{u},\bm{\Sigma}_\textrm{b},\bm{p}) \nonumber \\
		&=& \sum_{n=1}^{N_\textrm{t}} q_{n}  \sum_{t=1}^T \log \left(1+ p_{n}s_{n} \textrm{tr} \left[\bm{x}_{n}^\dag \bm{R}_n^{\frac{1}{2}} \bm{B}_\textrm{DFT} \bm{\Sigma}_\textrm{b} \bm{A}^{-1} \bm{\Sigma}_\textrm{b} \bm{B}_\textrm{DFT} ^\dag \bm{R}_n^{\frac{1}{2}} \bm{x}_{n} \right] \right) \nonumber\\
		\label{y1}
		&\xrightarrow{K \to \infty}& \sum_{n=1}^{N_\textrm{t}} q_{n}  \sum_{t=1}^T \log \left(1+ p_{n}s_{n} \textrm{tr} \left[\bm{\Sigma}_\textrm{b} \bm{B}_\textrm{DFT} ^\dag \bm{R}_n \bm{B}_\textrm{DFT} \bm{\Sigma}_\textrm{b} \bm{A}^{-1}  \right] \right) \\
		\label{y2}
		&\xrightarrow{K \to \infty}& \sum_{n=1}^{N_\textrm{t}} q_{n}  T \log \left(1+ p_{n}s_{n} \textrm{tr} \left[\bm{\Sigma}_\textrm{b} \bm{B}_\textrm{DFT} ^\dag \bm{R}_n \bm{B}_\textrm{DFT} \bm{\Sigma}_\textrm{b} 
		\left(\frac{1}{M}\sum_{j=1}^{n-1} \frac{p_j s_j \bm{\Sigma}_\textrm{b} \bm{B}_\textrm{DFT} ^\dag \bm{R}_j \bm{B}_\textrm{DFT} \bm{\Sigma}_\textrm{b} }{1+e_{n,j}} + \bm{I}\right)^{-1}  \right] \right), \nonumber\\
	\end{IEEEeqnarray}
	where $e_{n,i}$ is the unique solution of the following equations.
	\begin{equation}
	e_{n,i} = \textrm{tr} \left[\bm{\Sigma}_\textrm{b} \bm{B}_\textrm{DFT} ^\dag \bm{R}_i \bm{B}_\textrm{DFT} \bm{\Sigma}_\textrm{b} \left(\frac{1}{M}\sum_{j=1}^{n-1} \frac{p_j s_j \bm{\Sigma}_\textrm{b} \bm{B}_\textrm{DFT} ^\dag \bm{R}_j \bm{B}_\textrm{DFT} \bm{\Sigma}_\textrm{b}}{1+e_{n,j}} + \bm{I}\right)^{-1}  \right].
	\end{equation}
	The inequality \eqref{y1} is based on \cite[Lemma 14.2]{cou11}, and \eqref{y2} is based on the following lemma.
	\begin{lemma}
		\label{lm2}
		Let $\bm{B_N} = \bm{X}_N^\dag\bm{X}_N+\bm{S}_N$ with $\bm{S}_N \in \mathbb{C}^{N \times N}$ Hermitian nonnegative definite and $\bm{X}_N \in \mathbb{C}^{n \times N}$ random. The $i$th column $\bm{x}_i$ of $\bm{X}_N^\dag$ is $\bm{x}_i = \bm{R}_i^{\frac{1}{2}}\bm{y}_i$, where the entries of $\bm{y}_i \in \mathbb{C}^{r_i}$ are i.i.d. of zero mean, variance $1/N$ and have eighth-order moment of order $\mathcal{O}\left(\frac{1}{N^4}\right)$. The matrices $\bm{R}_i$'s are channel correlation matrices for each user, and $\bm{Q}_N \in \mathbb{C}^{N \times N}$ is deterministic. Assume $\limsup_{N \to \infty} \sup_{1 \le i \le N} \|\bm{R}_i\| < \infty$ and let $\bm{Q}_N$ have uniformly bounded spectral norm (with respect to $N$). Define
		\begin{equation}
		m_{\bm{B}_N,\bm{Q}_N}(z) = \frac{1}{N} \textrm{tr}\bm{Q}_N\left(\bm{B}_N-z\bm{I}_N\right)^{-1}.
		\end{equation}
		Then, for $z \in \mathbb{C} \backslash \mathbb{R}^+$, as $n$, $N$ grow large with ratios $\beta_{N,i}=N/r_i$ and $\beta=N/n$ such that $0<\liminf_N \beta_N \le \limsup_N \beta_N < \infty$ and $0<\liminf_N \beta_{N,i} \le \limsup_N \beta_{N,i} < \infty$, we have that 
		\begin{equation}
		m_{\bm{B}_N,\bm{Q}_N}(z) - m^o_{\bm{B}_N,\bm{Q}_N}(z) \to 0
		\end{equation}
		almost surely, with $m^o_{\bm{B}_N,\bm{Q}_N}(z)$ given by
		\begin{equation}
		m^o_{\bm{B}_N,\bm{Q}_N}(z) = \frac{1}{N} \textrm{tr}\bm{Q}_N \left(\frac{1}{N} \sum_{j=1}^n \frac{\bm{R}_j}{1+e_{N,j}(z)}+\bm{S}_N - z\bm{I}_N\right)^{-1}
		\end{equation}
		where the functions $e_{N,j}(z)$ form the unique solution of 
		\begin{equation}
		e_{N,i}(z) = \frac{1}{N} \textrm{tr}\bm{R}_i \left(\frac{1}{N} \sum_{j=1}^n \frac{\bm{R}_j}{1+e_{N,j}(z)}+\bm{S}_N - z\bm{I}_N\right)^{-1}
		\end{equation}
	\end{lemma}
	
	\section{Proof of the Convergence of the BCU-based Algorithm}
	\label{app3}
	First, the proof of the iterative water filling approach in the user scheduling part is given. Consider the user scheduling problem with beam selection fixed, i.e.,
	\begin{flalign}
		\label{P5}
		\textbf{P5:}&&\mathop{\textrm{maximize}}\limits_{\bm{w}}  \,\,&  \sum_{n=1}^{N_\textrm{t}} q_{n}  \log \left(1+ \mathsf{f}_n(\bm{w}) \right) &&\\
		&&\textrm{s.t.,}\,\,  & \sum_{n=1}^{N_\textrm{t}} w_n \le P,&&
	\end{flalign}
	Denote 
	\begin{equation}
	\mathsf{f}_n(\bm{w})= \mathsf{f}_n(w_1,w_2,...,w_{n}) = w_{n} \textrm{tr} \left[\bm{\Sigma}_\textrm{b} \bm{B}_\textrm{DFT} ^\dag \bm{R}_n \bm{B}_\textrm{DFT} \bm{\Sigma}_\textrm{b} \left(\frac{1}{M}\sum_{j=1}^{n-1} \frac{w_j  \bm{R}_j}{1+e_{n,j}} + \bm{I}\right)^{-1}  \right].
	\end{equation}
	The key to the proof is to construct the equivalent optimization problem as stated below.
	\begin{flalign}
		\label{P6}
		\textbf{P6:}&&\mathop{\textrm{maximize}}\limits_{\bm{w}(m),\, 0\le m \le N_\textrm{t}-1}  \,\,&  \frac{1}{N_\textrm{t}}\sum_{m=0}^{N_\textrm{t}-1}\sum_{n=1}^{N_\textrm{t}} q_{n}  \log \left(1+ \mathsf{f}_n(w_1([m+1]_{N_\textrm{t}}),w_2([m+2]_{N_\textrm{t}}),...,w_n([m+n]_{N_\textrm{t}}))\right) &&\\
		&&\textrm{s.t.,}\,\,  & \sum_{n=1}^{N_\textrm{t}} w_n(m) \le P,\,\forall m. &&
	\end{flalign}
	The reason that \textbf{P5} and \textbf{P6} are equivalent is straightforward due to the Shur-concavity of the objective function of \textbf{P6} \cite{marshall79}. Therefore, the solution of \textbf{P6} is obtained at the point which satisfies
	\begin{equation}
	\bm{w}(m) = \bm{w},\,\forall m.
	\end{equation}
	Since \textbf{P6} is concave in $\bm{w}(m)$, the BCU technique which cyclically optimizes $\bm{w}(m)$ with others fixed is guaranteed to converge to the global optimum, which yields the same procedure as in Algorithm \ref{alg:bcu} with some minor mathematical manipulations \cite{kobayashi06}. Therefore, we conclude that the iterative water filling approach adopted in Algorithm \ref{alg:bcu} converges to the optimum in the user scheduling step.
	
	Next, it will be shown that the BCU technique which cyclically update user scheduling and beam selection converges to the optimum. Based on \cite{warga63}, it is sufficient to check if the problem satisfies the following two conditions:
	\begin{itemize}
		\item 
		The objective function, denoted by $\phi(\bm{x})$, is continuously differentiable in some neighborhood of every stationary point of $\phi(\bm{x})$.
		\item
		For every $k$, $1 \le k \le n$, $\phi(\bm{x})$ is a strictly concave function of $x_k$, the other points $x_j$, $j \neq k$, being arbitrarily chosen in their respective domains.
	\end{itemize}
	
	The above two conditions are easily met in this problem since $\mathcal{D}(\bm{Q},\bm{R}_1,...,\bm{R}_{N_\textrm{t}},\bm{I}_{N_\textrm{t}},\bm{\Sigma}_\textrm{b},\bm{w})$ is continuously differentiable in the whole domain and concave in $\bm{\Sigma}_\textrm{b}$ and $\bm{w}$, respectively. Therefore, the proposed BCU-based scheduling scheme is guaranteed to converge to the global optimum.
	
	\bibliographystyle{ieeetr}
	\bibliography{15}

\begin{thebibliography}{10}

\bibitem{Marzetta10}
T.~Marzetta, ``Noncooperative cellular wireless with unlimited numbers of base
  station antennas,'' {\em IEEE Trans. Wireless Commun.}, vol.~9,
  pp.~3590--3600, Nov 2010.

\bibitem{Rusek12}
F.~Rusek, A.~Lozano, and N.~Jindal, ``Mutual information of {IID} complex
  {G}aussian signals on block {R}ayleigh-faded channels,'' {\em IEEE Trans.
  Inform. Theory}, vol.~58, pp.~331--340, Jan. 2012.

\bibitem{larsson14}
E.~G. Larsson, O.~Edfors, F.~Tufvesson, and T.~L. Marzetta, ``Massive {MIMO}
  for next generation wireless systems,'' {\em IEEE Commun. Mag.}, vol.~52,
  pp.~186--195, Feb. 2014.

\bibitem{heath16}
R.~W. Heath, N.~González-Prelcic, S.~Rangan, W.~Roh, and A.~M. Sayeed, ``An
  overview of signal processing techniques for millimeter wave {MIMO}
  systems,'' {\em IEEE J. Sel. Top. Signal Process.}, vol.~10, pp.~436--453,
  Apr. 2016.

\bibitem{han15}
S.~Han, C.~l.~I, Z.~Xu, and C.~Rowell, ``Large-scale antenna systems with
  hybrid analog and digital beamforming for millimeter wave {5G},'' {\em IEEE
  Commun. Mag.}, vol.~53, pp.~186--194, Jan. 2015.

\bibitem{jiang_icc17}
Z.~Jiang, S.~Zhou, and Z.~Niu, ``Antenna-beam spatial transformation in {C-RAN}
  with large antenna arrays,'' in {\em IEEE Int. Conf. Commun. (ICC)
  Workshops}, May 2017.

\bibitem{alkh14}
A.~Alkhateeb, O.~E. Ayach, G.~Leus, and R.~W. Heath, ``Channel estimation and
  hybrid precoding for millimeter wave cellular systems,'' {\em IEEE J. Sel.
  Top. Signal Process.}, vol.~8, pp.~831--846, Oct 2014.

\bibitem{molisch16}
A.~F. Molisch, V.~V. Ratnam, S.~Han, Z.~Li, S.~L.~H. Nguyen, L.~Li, and
  K.~Haneda, ``Hybrid beamforming for massive {MIMO}: A survey,'' {\em IEEE
  Commun. Mag.}, vol.~55, pp.~134--141, Sep 2017.

\bibitem{gao16}
X.~Gao, L.~Dai, S.~Han, C.~L. I, and R.~W. Heath, ``Energy-efficient hybrid
  analog and digital precoding for mmwave {MIMO} systems with large antenna
  arrays,'' {\em IEEE J. Select. Areas Commun.}, vol.~34, pp.~998--1009, Apr.
  2016.

\bibitem{brady13}
J.~Brady, N.~Behdad, and A.~M. Sayeed, ``Beamspace {MIMO} for millimeter-wave
  communications: System architecture, modeling, analysis, and measurements,''
  {\em IEEE Trans. Antennas and Propagat.}, vol.~61, pp.~3814--3827, Jul. 2013.

\bibitem{Jiang14}
Z.~Jiang, A.~Molisch, G.~Caire, and Z.~Niu, ``Achievable rates of {FDD} massive
  {MIMO} systems with spatial channel correlation,'' {\em IEEE Trans. Wireless
  Commun.}, vol.~14, pp.~2868--2882, May 2015.

\bibitem{7}
Z.~Jiang, S.~Zhou, and Z.~Niu, ``On dimensionality loss in {FDD} massive {MIMO}
  systems,'' in {\em IEEE Conf. Wireless Commun. Netw. (WCNC)}, pp.~399--404,
  Mar 2015.

\bibitem{jiang17}
Z.~Jiang, S.~Zhou, R.~Deng, Z.~Niu, and S.~Cao, ``Pilot-data superposition for
  beam-based {FDD} massive {MIMO} downlinks,'' {\em IEEE Commun. Letters},
  vol.~21, pp.~1357--1360, Jun 2017.

\bibitem{Adhikary13}
A.~Adhikary, J.~Nam, J.-Y. Ahn, and G.~Caire, ``Joint spatial division and
  multiplexing: The large-scale array regime,'' {\em IEEE Trans. Inform.
  Theory}, vol.~59, pp.~6441--6463, Oct. 2013.

\bibitem{molisch04}
A.~F. Molisch and X.~Zhang, ``{FFT}-based hybrid antenna selection schemes for
  spatially correlated {MIMO} channels,'' {\em IEEE Commun. Letters}, vol.~8,
  pp.~36--38, Jan 2004.

\bibitem{Ayach14}
O.~E. Ayach, S.~Rajagopal, S.~Abu-Surra, Z.~Pi, and R.~W. Heath, ``Spatially
  sparse precoding in millimeter wave {MIMO} systems,'' {\em IEEE Trans.
  Wireless Commun.}, vol.~13, pp.~1499--1513, Mar. 2014.

\bibitem{Yoo2006}
T.~Yoo and A.~Goldsmith, ``On the optimality of multiantenna broadcast
  scheduling using zero-forcing beamforming,'' {\em IEEE J. Select. Areas
  Commun.}, vol.~24, pp.~528--541, Mar. 2006.

\bibitem{shirani10}
H.~Shirani-Mehr, G.~Caire, and M.~J. Neely, ``{MIMO} downlink scheduling with
  non-perfect channel state knowledge,'' {\em IEEE Trans. Commun.}, vol.~58,
  pp.~2055--2066, July 2010.

\bibitem{kim04}
H.~Kim, K.~Kim, Y.~Han, and S.~Yun, ``A proportional fair scheduling for
  multicarrier transmission systems,'' in {\em IEEE Veh. Tech. Conf.
  (VTC-Fall)}, vol.~1, pp.~409--413, Sep 2004.

\bibitem{bas17}
C.~U. Bas, R.~Wang, D.~Psychoudakis, T.~Henige, R.~Monroe, J.~Park, J.~Zhang,
  and A.~F. Molisch, ``A real-time millimeter-wave phased array {MIMO} channel
  sounder,'' {\em arXiv preprint arXiv:1703.05271}, 2017.

\bibitem{gao16_bs}
X.~Gao, L.~Dai, Z.~Chen, Z.~Wang, and Z.~Zhang, ``Near-optimal beam selection
  for beamspace mmwave massive {MIMO} systems,'' {\em IEEE Commun. Letters},
  vol.~20, pp.~1054--1057, May 2016.

\bibitem{ama15}
P.~V. Amadori and C.~Masouros, ``Low {RF}-complexity millimeter-wave
  beamspace-{MIMO} systems by beam selection,'' {\em IEEE Trans. Commun.},
  vol.~63, pp.~2212--2223, Jun 2015.

\bibitem{liang01}
Y.-C. Liang and F.~P.~S. Chin, ``Downlink channel covariance matrix ({DCCM})
  estimation and its applications in wireless {DS-CDMA} systems,'' {\em IEEE J.
  Select. Areas Commun.}, vol.~19, pp.~222--232, Feb 2001.

\bibitem{Adhikary14}
A.~Adhikary, E.~Al~Safadi, M.~Samimi, R.~Wang, G.~Caire, T.~Rappaport, and
  A.~Molisch, ``Joint spatial division and multiplexing for mm-wave channels,''
  {\em IEEE J. Select. Areas Commun.}, vol.~32, pp.~1239--1255, Jun. 2014.

\bibitem{che15}
A.~Checko, H.~L. Christiansen, Y.~Yan, L.~Scolari, G.~Kardaras, M.~S. Berger,
  and L.~Dittmann, ``Cloud {RAN} for mobile networks -- {A} technology
  overview,'' {\em IEEE Commun. Surveys Tuts.}, vol.~17, pp.~405--426,
  Firstquarter 2015.

\bibitem{gao17_sd}
X.~Gao, L.~Dai, S.~Han, C.~L. I, and X.~Wang, ``Reliable beamspace channel
  estimation for millimeter-wave massive {MIMO} systems with lens antenna
  array,'' {\em IEEE Trans. Wireless Commun.}, vol.~16, pp.~6010--6021, Sep
  2017.

\bibitem{jiang17_apcc}
Z.~Jiang, S.~Zhou, and Z.~Niu, ``A block coordinated update method for
  beam-based massive {MIMO} downlink scheduling based on statistical {CSI},''
  in {\em Asia-Pacific Conference on Communications (APCC)}, Dec 2017.

\bibitem{wang16}
J.~Wang, H.~Zhu, L.~Dai, N.~J. Gomes, and J.~Wang, ``Low-complexity beam
  allocation for switched-beam based multiuser massive {MIMO} systems,'' {\em
  IEEE Trans. Wireless Commun.}, vol.~15, pp.~8236--8248, Dec 2016.

\bibitem{sun15}
C.~Sun, X.~Gao, S.~Jin, M.~Matthaiou, Z.~Ding, and C.~Xiao, ``Beam division
  multiple access transmission for massive {MIMO} communications,'' {\em IEEE
  Trans. Commun.}, vol.~63, pp.~2170--2184, Jun 2015.

\bibitem{you17}
L.~You, X.~Gao, G.~Y. Li, X.~G. Xia, and N.~Ma, ``{BDMA} for
  millimeter-wave/terahertz massive {MIMO} transmission with per-beam
  synchronization,'' {\em IEEE J. Select. Areas Commun.}, vol.~35,
  pp.~1550--1563, Jul 2017.

\bibitem{molisch04mag}
A.~F. Molisch and M.~Z. Win, ``{MIMO} systems with antenna selection,'' {\em
  IEEE Microwave Mag.}, vol.~5, pp.~46--56, Mar 2004.

\bibitem{neely10}
M.~J. Neely, ``Stochastic network optimization with application to
  communication and queueing systems,'' {\em Synthesis Lectures on
  Communication Networks}, vol.~3, no.~1, pp.~1--211, 2010.

\bibitem{edelman05}
A.~Edelman and N.~R. Rao, ``Random matrix theory,'' {\em Acta Numerica},
  vol.~14, pp.~233--297, 2005.

\bibitem{warga63}
J.~Warga, ``Minimizing certain convex functions,'' {\em Journal of the Society
  for Industrial and Applied Mathematics}, vol.~11, no.~3, pp.~588--593, 1963.

\bibitem{nem78}
G.~L. Nemhauser, L.~A. Wolsey, and M.~L. Fisher, ``An analysis of
  approximations for maximizing submodular set functions---{I},'' {\em
  Mathematical Programming}, vol.~14, no.~1, pp.~265--294, 1978.

\bibitem{molisch04_gscm}
A.~Molisch, ``A generic model for {MIMO} wireless propagation channels in
  macro- and microcells,'' {\em IEEE Trans. Signal Process.}, vol.~52,
  pp.~61--71, Jan. 2004.

\bibitem{kush04}
H.~J. Kushner and P.~A. Whiting, ``Convergence of proportional-fair sharing
  algorithms under general conditions,'' {\em IEEE Trans. Wireless Commun.},
  vol.~3, pp.~1250--1259, Jul 2004.

\bibitem{neely08}
M.~J. Neely, E.~Modiano, and C.~P. Li, ``Fairness and optimal stochastic
  control for heterogeneous networks,'' {\em IEEE/ACM Trans. Netw.}, vol.~16,
  pp.~396--409, Apr 2008.

\bibitem{Weingarten06}
H.~Weingarten, Y.~Steinberg, and S.~Shamai, ``The capacity region of the
  {G}aussian multiple-input multiple-output broadcast channel,'' {\em IEEE
  Trans. Inform. Theory}, vol.~52, pp.~3936--3964, Sep. 2006.

\bibitem{Wei06}
W.~Yu, ``Uplink-downlink duality via minimax duality,'' {\em IEEE Trans.
  Inform. Theory}, vol.~52, pp.~361--374, Feb. 2006.

\bibitem{Jiang12}
Z.~Jiang, S.~Zhou, and Z.~Niu, ``Capacity bounds of downlink network {MIMO}
  systems with inter-cluster interference,'' in {\em IEEE Global Commun. Conf.
  (GLOBECOM)}, pp.~4612--4617, Dec. 2012.

\bibitem{Jang02}
J.~W. Lee, R.~R. Mazumdar, and N.~B. Shroff, ``Downlink power allocation for
  multi-class {CDMA} wireless networks,'' in {\em IEEE INFOCOM}, vol.~3,
  pp.~1480--1489 vol.3, 2002.

\bibitem{geo06}
L.~Georgiadis, M.~J. Neely, and L.~Tassiulas, {\em Resource allocation and
  cross-layer control in wireless networks}.
\newblock Now Publishers Inc, 2006.

\bibitem{karp72}
R.~M. Karp, ``Reducibility among combinatorial problems,'' in {\em Complexity
  of computer computations}, pp.~85--103, Springer, 1972.

\bibitem{Wei04}
W.~Yu and J.~Cioffi, ``Sum capacity of {Gaussian} vector broadcast channels,''
  {\em IEEE Trans. Inform. Theory}, vol.~50, pp.~1875--1892, Sep. 2004.

\bibitem{nai10}
S.~E. Nai, W.~Ser, Z.~L. Yu, and H.~Chen, ``Beampattern synthesis for linear
  and planar arrays with antenna selection by convex optimization,'' {\em IEEE
  Trans. Antennas and Propag.}, vol.~58, pp.~3923--3930, Dec 2010.

\bibitem{dua06}
A.~Dua, K.~Medepalli, and A.~J. Paulraj, ``Receive antenna selection in {MIMO}
  systems using convex optimization,'' {\em IEEE Trans. Wireless Commun.},
  vol.~5, pp.~2353--2357, Sep. 2006.

\bibitem{jindal05}
N.~Jindal, W.~Rhee, S.~Vishwanath, S.~A. Jafar, and A.~Goldsmith, ``Sum power
  iterative water-filling for multi-antenna {Gaussian} broadcast channels,''
  {\em IEEE Trans. Inform. Theory}, vol.~51, pp.~1570--1580, Apr 2005.

\bibitem{Tseng01}
P.~Tseng, ``Convergence of a block coordinate descent method for
  nondifferentiable minimization,'' {\em Journal of Optimization Theory and
  Applications}, vol.~109, no.~3, pp.~475--494, 2001.

\bibitem{kim16}
R.~Kim, H.~Lim, and B.~Krishnamachari, ``Prefetching-based data dissemination
  in vehicular cloud systems,'' {\em IEEE Trans. Veh. Tech.}, vol.~65,
  pp.~292--306, Jan 2016.

\bibitem{rap13}
T.~S. Rappaport, S.~Sun, R.~Mayzus, H.~Zhao, Y.~Azar, K.~Wang, G.~N. Wong,
  J.~K. Schulz, M.~Samimi, and F.~Gutierrez, ``Millimeter wave mobile
  communications for {5G} cellular: It will work!,'' {\em IEEE Access}, vol.~1,
  pp.~335--349, 2013.

\bibitem{peel05}
C.~Peel, B.~Hochwald, and A.~Swindlehurst, ``A vector-perturbation technique
  for near-capacity multiantenna multiuser communication-part {I}: Channel
  inversion and regularization,'' {\em IEEE Trans Commun.}, vol.~53,
  pp.~195--202, Jan. 2005.

\bibitem{hoch04}
B.~M. Hochwald, T.~L. Marzetta, and V.~Tarokh, ``Multiple-antenna channel
  hardening and its implications for rate feedback and scheduling,'' {\em IEEE
  Trans. Inform. Theory}, vol.~50, pp.~1893--1909, Sep 2004.

\bibitem{4}
Z.~Jiang, S.~Zhou, and Z.~Niu, ``Dynamic channel acquisition in {MU-MIMO},''
  {\em IEEE Trans. Commun.}, vol.~62, pp.~4336--4348, Dec. 2014.

\bibitem{Wagner12}
S.~Wagner, R.~Couillet, M.~Debbah, and D.~T.~M. Slock, ``Large system analysis
  of linear precoding in correlated {MISO} broadcast channels under limited
  feedback,'' {\em IEEE Trans. Inform. Theory}, vol.~58, pp.~4509--4537, Jul.
  2012.

\bibitem{cou11}
R.~Couillet and M.~Debbah, {\em Random matrix methods for wireless
  communications}.
\newblock Cambridge University Press, 2011.

\bibitem{marshall79}
A.~W. Marshall, I.~Olkin, and B.~C. Arnold, {\em Inequalities: theory of
  majorization and its applications}, vol.~143.
\newblock Springer, 1979.

\bibitem{kobayashi06}
M.~Kobayashi and G.~Caire, ``An iterative water-filling algorithm for maximum
  weighted sum-rate of {Gaussian MIMO-BC},'' {\em IEEE J. Select. Areas
  Commun.}, vol.~24, pp.~1640--1646, Aug 2006.

\end{thebibliography}
\end{document}